\documentclass[a4paper,11pt]{article}
\usepackage{jcappub} 
\usepackage{lineno}
\usepackage{graphicx}
\usepackage{amsmath}
\usepackage{amssymb}
\usepackage{comment}
\usepackage{xcolor}
\usepackage{subfigure, rotating, bm, array}

\pdfoutput=1 
\usepackage[T1]{fontenc} 

\title{Sub-Horizon and Quasi-Static Approximations in Interacting Dark Sector Models Involving Scalar Field Dark Energy
}

\author[a]{Priyanka Saha,}
\author[b]{Sharvari Nadkarni-Ghosh,}
\author[a]{Kaushik bhattacharya}
\affiliation[a]{Department of Physics, Indian Institute of Technology, Kanpur,\\
Kanpur-208016, India}
\affiliation[b]{Department of Space, Planetary \& Astronomical Sciences \& Engineering,\\
Indian Institute of Technology, Kanpur,\\
Kanpur-208016, India}

\emailAdd{priyankas21@iitk.ac.in}
\emailAdd{sharvari@iitk.ac.in}
\emailAdd{kaushikb@iitk.ac.in}

\abstract{
We investigate the validity of the sub-horizon approximation (SHA) and quasi-static approximation (QSA) for linear matter perturbations in interacting scalar-field dark-energy models. We consider both quintessence and phantom scalar fields and derived their full relativistic perturbation evolution to compare with the corresponding SHA and QSA solutions. The accuracy of the approximations is quantified through the matter growth factor, growth rate, and linear matter power spectrum over a range of comoving wavenumbers and interaction parameters. We find that, while the SHA reproduces the full evolution with percent-level accuracy for non-interacting and weakly interacting models on sufficiently sub-horizon scales, its accuracy progressively deteriorates with increasing interaction strength and towards larger scales.  Moreover, unlike in \(\Lambda\)CDM and non-interacting scalar-field models, applying the SHA to interacting models does not fully recover the standard Newtonian fluid equations, as additional interaction-dependent terms remain. The QSA generally provides a closer description of the full relativistic evolution than the SHA over the scales considered. By comparing the scalar-field model with $\Lambda$CDM using the full relativistic equations and, separately, using the SHA system, we evaluated whether the SHA is capable of capturing the full extent of the physical differences introduced by the alternative model. We find that the difference between these two comparisons becomes increasingly important for stronger interactions and smaller wavenumbers, indicating a growing contribution from the perturbation dynamics. Finally, we demonstrate that the resulting differences can reach several percent on scales accessible to large-volume hydrodynamical simulations, implying that the validity of the SHA cannot be assumed solely from the sub-horizon condition. Our results highlight the need to benchmark the approximation over the relevant interaction parameter and wavenumber ranges before employing interacting-dark-energy models in large-volume structure-formation simulations and precision comparisons with large-scale-structure observations.}

\begin{document}
\maketitle
\flushbottom
\section{Introduction}
Many upcoming Large Scale Structure (LSS) surveys such as Euclid\footnote{http://sci.esa.int/euclid/.}, DESI\footnote{https://www.desi.lbl.gov/}, Vera C. Rubin Observatory\footnote{https://www.lsst.org}, SPHEREx\footnote{https://spherex.caltech.edu/} and the Nancy Grace Roman Telescope \footnote{https://science.nasa.gov/mission/roman-space-telescope/} all aim to map the large scale matter distribution on scales nearing one gigaparsec. One of the primary aims of these surveys is to constrain cosmological parameters. However, this exercise relies on accurate theoretical predictions of the relevant observables. These observables are the outputs of N-body codes or perturbative techniques which solve the equations that govern the growth of fluctuations. 

The matter distribution that we observe today grew from tiny inhomogeneities imprinted in the cosmological fluid at very early epochs. At very early epochs, radiation plays an important role in the growth of matter fluctuations and Einstein's equations of general relativity (GR) coupled with the Boltzmann equation for the distribution of various species are the equations that govern this growth. Fortunately, at these epochs the fluctuations are small and it suffices to consider the linearized system; linear in the perturbation variables that quantify departures from the Friedmann-Lemaitre-Robertson-Walker (FLRW) metric and uniform stress-energy tensor. This complete linear system is still fairly complicated due to the large number of Fourier modes and is solved by transfer function codes, such as CAMB\cite{Lewis_2000,lewis_2026}, CLASS \cite{lesgourgues2011,Diego_Blas_2011} or PyCOSMO \cite{refregier2017}. As the perturbations enter matter domination, they start to grow under gravity and the linear approximation breaks down. The full non-linear Einstein's equations of GR are far too complex and other simplifications need to be invoked to make further progress in the non-linear regime. \footnote{In a non-cosmological context, usually one invokes the weak field limit to recover Newtonian dynamics; this means assuming that the dimensionless potentials are small and/or that the velocities are non-relativistic.} One simplifying assumption is the Sub-Horizon Approximation (SHA), which restricts the system to perturbations on scales much smaller than the horizon. In this limit, the GR equations reduce to the usual equations of Newtonian Hydrodynamics - the continuity equation and the Euler equation for density and velocity coupled with Poisson's equation for the gravitational force. A decade ago, this approximation held true for most observational probes (e.g, SDSS-I,  2DFGRS \footnote{http://www.2dfgrs.net/}) and N-body codes (e.g., see review paper \cite{Angulo_2022}) or perturbative techniques (e.g.,see review paper \cite{bernardeau_2002}) solved this coupled system with an accuracy more than the survey systematics. These observations along with the CMB maps, validated and choose the $\Lambda$CDM model as the most preferred one. However, recent observations such as the Hubble or $\sigma_8$ tension (for e.g., ~\cite{Macaulay_2013,Di_Valentino_2021}) have re-emphasized the need to look beyond the $\Lambda$CDM. Many alternative theories have been proposed, including models involving quintessence and phantom fields, in which an additional scalar field 
may couple to the dark matter sector, as well as theories which modify the Einstein-Hilbert action\footnote{For certain class of models, these two approaches are equivalent through metric transformations (see for e.g., \cite{joyce_dark_2016, wettrich_2015, odinstov2025}) }
\cite{Peebles_2003,Tsujikawa_2013,sotiriou_fr_2010,defelice_fR_2010,clifton_modified_2012,Nojiri_2017}.

If no additional characteristic timescale is present, the quasi-static approximation (QSA) may be interpreted as an ordering in which the dimensionless temporal derivatives of the perturbations are small compared with the corresponding dimensionless spatial-gradient terms. Complementing this, the Sub-Horizon Approximation (SHA) imposes a spatial hierarchy, assuming that the physical wavelengths of the perturbations are well within the Hubble radius ($k \gg \mathcal{H}$), allowing terms suppressed by $\mathcal{H}/k$ to be neglected. Both orderings are naturally realized in \(\Lambda\)CDM on sufficiently small scales, particularly during matter domination. In alternative cosmological models, however, additional dynamical degrees of freedom can introduce characteristic timescales independent of the Hubble scale, including rapidly varying or oscillatory modes \cite{Nadkarni_Ghosh_2022, Nadkarni-Ghosh_2026}, and hence the validity of the QSA and SHA is no longer guaranteed in these cases.
It is therefore important to assess the validity of the SHA and QSA independently for these models. Given the range of scales probed by present and upcoming surveys, quantifying the accuracy of these approximations in both \(\Lambda\)CDM and alternative cosmologies is of practical importance. Although their widespread use is partly motivated by the computational complexity of nonlinear modelling, establishing their domain of validity in the linear regime provides a necessary baseline for assessing their subsequent application to nonlinear structure formation.

Early investigations of dark energy described by a scalar field implicitly ignored the rapid oscillations in the scalar field (for example, \cite{Amendola_2004,Koivisto_2005}). Such rapid oscillations break the QSA but ignoring them was given observational justification by Silvestri {\it et al} \cite{Hojjati_2012,silvestri_practical_2013} who argued that these oscillations are not detectable and can be accounted for by appropriate theoretical priors in parameter estimation.  In contrast, Bamba {\it et al} \cite{Bamba:2011ih}, examined cosmological perturbations in $k$-essence dark energy models with and without imposing the SHA. Their analysis showed that the non-canonical kinetic structure of the scalar field can generate rapidly oscillating modes in the matter density perturbations whose behaviour is governed by the effective sound speed of the $k$-essence field. They also argue that for a wavenumber $k \sim 0.01 h^{-1} {\rm Mpc}$, these oscillations may be detectable, providing constraints on these kinds of models. Llinares and Mota \cite{Llinares:2013qbh} argue that oscillations in scalar-tensor theories should leave imprints on galactic and cluster scales which should be detectable in $X$-rays. More recently, Mirpoorian {\it et al} \cite{Mirpoorian_2023} mapped out the range of validity of the QSA in the symmetron model as an example scalar-tensor theory and validated the use of QSA in transfer function generating variants such as MGCAMB \cite{Zhao_2009,Hojjati_2011,Zucca_2019} or MGCLASS \cite{Baker_2015,Sakr_2022}. Other authors have also used SHA and QSA to simplify the perturbation equations in the context of Horndeski models or Effective Fluid approach so that they can be easily incorporated into Boltzmann codes (e.g., \cite{Arjona_2019, Cardona_2022,Cataneo_2024}); see Nesseris \cite{Nesseris_2022} for a comprehensive review.  The presence of oscillations and the potential breakdown of QSA has also been discussed in the context of $f(R)$ theories. De la Cruz-Dombriz \cite{delaCruz-Dombriz:2008ium} obtained fourth order temporal differential equations for the two metric potentials and showed that their time evolution does not satisfy the QSA and can breakdown on sub-horizon scales. Sawicki and Bellini \cite{Sawicki:2015zya} showed that the applicability of QSA is controlled not by the cosmological horizon, but by the sound horizon of the dark energy fluid. It is also known that there are high frequency oscillations even at the background level in $f(R)$ models such as the one by Hu and Sawicki \cite{hu_models_2007,song_large_2007,Nadkarni_Ghosh_2022}. Noller {\it et al} \cite{Noller:2013wca} more systematically quantified the errors  introduced by the QSA as a function of parameters of the theory such as the dark energy density parameter and the epoch that the approximation is turned on. Pace {\it et al} \cite{pace_comparison_2021} investigated the implementation of QSA for Horndeski gravity models at three different levels - the field equations, effective equations for scalar potentials and the equation of state approach - to conclude that all three approaches agree on observationally relevant scales justifying the widespread use of this approximation. Recently, Orjuela-Quintana and Nesseris \cite{Orjuela-Quintana_2023} provided a 
transparent theoretical framework, by defining eight parameters that quantify the QSA and SHA by capturing higher order time derivatives and examined the range of validity of the approximations in $f(R)$ theories. A somewhat similar investigation for testing the validity of the SHA was undertaken by Amin {\it et al} \cite{Amin:2007wi} wherein they formally expanded the relation between the metric and the density fields in Fourier space in terms of powers of the parameter ($ck/aH$) which quantifies the SHA. 
This framework provides a controlled way to go beyond the standard Poisson equation and captures scale-dependent corrections arising in modified gravity theories.

On the non-linear front, there have been efforts to see deviations from QSA and SHA in numerical simulations.  Llinares and collaborators \cite{Llinares_2014,Llinares_2014b} modified the publicly available code RAMSES \cite{Teyssier_2002} and explicitly checked the QSA in the symmetron model. They find that while, on smaller scales there is an effect on the matter power spectrum calculated after filtering local distributions the global power spectrum is largely unaffected. Bose {\it et al} \cite{Bose_2015} checked the QSA in the Hu-Sawicki $f(R)$  theory to conclude that the time derivatives of the effective scalar field can be neglected. Winther {\it et al} \cite{Winther_2017} conducted a more extensive comparison of different modified gravity N-body codes applied to different types of models and came to a similar conclusion that the QSA is a good approximation. 
Most of these investigations that relax the QSA assume that the potentials corresponding to the scalar fields are time-dependent, but they still treat the potentials corresponding to the metric perturbations as quasi-static. 
The validity of N-body simulations on very large scales has also been discussed in the context of the $\Lambda$CDM model in Refs.~\cite{Adamek_2014,Rigopoulos_2015,Fidler_2015,Adamek_2016,Adamek_2016b}. While the authors of these papers have not implemented scalar field or modified gravity models, they have implicitly checked the validity of the SHA at large scales.

In the present work, our primary focus is on validation of the SHA and the QSA in interacting dark sector models where the dynamical dark energy candidate is a scalar field and we compare its predictions with those of the corresponding non-interacting scalar field dark energy model and the standard $\Lambda$CDM model. 
The importance of testing these approximations becomes particularly pronounced in interacting dark-sector models \cite{Saha:2025bjw,Saha:2025mwm,Saha:2024irh,Saha:2024xbg,Hussain:2023kwk,Bhattacharya:2022wzu,Chatterjee:2021ijw}. Unlike minimally coupled scalar-field dark energy, a direct interaction between dark matter and the scalar field modifies both the background evolution and the coupled perturbation dynamics, potentially introducing additional scales and time dependence for which the usual assumptions underlying the SHA and QSA need not remain valid. Nevertheless, the accuracy of these widely used approximations has received comparatively little attention in such interacting scenarios. In this work, we therefore test the SHA and QSA directly and independently against the full linear perturbation system for an interacting scalar-field dark-energy model. This provides a quantitative assessment of whether approximations justified in \(\Lambda\)CDM and uncoupled dark-energy models can be reliably carried over to an interacting dark sector, and identifies the scales and interaction strengths for which their predictions begin to depart from the full dynamics. In all the cases studied in the present paper, the dark matter component is assumed to be identical and is modeled as a cold, pressureless fluid with energy density $\rho_{m}$. The distinction between the models arises solely from the description of the dark energy sector.

In the interacting and non-interacting scalar field models, the late-time accelerated expansion of the Universe is driven by a dynamical scalar field $\phi$. The scalar field evolves with cosmic time, and its energy density $\rho_{\phi}$ and pressure $P_{\phi}$ are determined by the combined contributions of its kinetic and potential energy. As a result, the dark energy density evolves with time, giving rise to a time-dependent equation of state, $\omega_\phi \equiv \frac{P_\phi}{\rho_\phi}$. In the standard $\Lambda$CDM model, the dark energy is described by the cosmological constant $\Lambda$, corresponding to a constant dark energy density throughout cosmic evolution. Since our analysis starts in the matter-dominated era, where the radiation density is negligible compared to matter, we neglect the radiation and other subdominant components. Therefore, in all our models we consider a Universe consisting only of dark matter and dark energy. Throughout this work, background quantities are denoted by an overbar to distinguish them from the corresponding total quantities. We also adopt the natural unit system $8\pi G=c=1$. 

The present paper is organized in the following way. The next section sets the notation and conventions used throughout the paper. Section \ref{Approximations} presents the definition of the SHA and QSA. The next section \ref{bcp} gives a description of the background cosmological model used in this paper. The next section \ref{perturb} sets the cosmological perturbation equations in the interacting, as well as the non-interacting and $\Lambda$CDM models. The methodology in which we study the approximations in the various cases is discussed in section \ref{method}. The results of our analysis is presented in section \ref{result} and the paper concludes with a detailed discussion in section \ref{concl}. 
\section{Action and the energy momentum tensor}

The large-scale geometry of the Universe is well described by the spatially flat FLRW metric, which assumes homogeneity and isotropy on cosmological scales. In comoving coordinates and conformal time $\tau$, the background metric is given by,
\begin{equation}\label{backgroundmetric}
    ds^{2} = a^{2}\left(\tau\right) \left(-d\tau^{2} + \delta_{ij}\,dx^{i}dx^{j}\right),
\end{equation}
where $a\left(\tau\right)$ denotes the scale factor.
Now, the general action for an interacting two-fluid system can be written as \cite{Boehmer:2015kta, Saha:2024xbg}:
\begin{eqnarray}\label{action}
\mathcal{S}=\int d^4x~\left(\mathcal{L}_{\text{grav}}+\mathcal{L}_{\text{m}}+\mathcal{L}_{\phi}+\mathcal{L}_{\text{int}}\right)\,\, ,
\end{eqnarray}
where the gravitational sector is given by the standard Einstein–Hilbert Lagrangian:
\begin{eqnarray}\label{lagrangian-einstein_hilbert}
\mathcal{L}_{\text{grav}}=\frac{\sqrt[•]{-g}(R-2\Lambda)}{2}\,\, ,
\end{eqnarray}
here $g$ is the determinant of the metric tensor $g_{\mu\nu}$ and $R$ is the Ricci scalar. $\Lambda$ is set to be zero for non-$\Lambda$CDM models.
The Lagrangian for the relativistic matter fluid is given by,
\begin{eqnarray}\label{lagrangian-matter}
\mathcal{L}_{\text{m}}=-\sqrt[]{-g}\rho_{m}\left(n,s\right)+J^{\mu}\left(\varkappa_{,\mu}+s\theta_{,\mu}+\beta_{A}\alpha^{A}_{,\mu}\right)\,\, ,
\end{eqnarray}
where $\rho_{m}$ is the energy density of the matter, $\varkappa$, $\theta$, and $\beta_{A}$ are all Lagrange multipliers with A taking the values 1, 2, 3, and $\alpha^{A}$ are the Lagrangian coordinates of the fluid. 
The vector density or the current density of particle number $J^{\mu}$ is related to $n$ as
$$J^{\mu}=\sqrt[]{-g}nu^{\mu}_{(m)},~ |J|=\sqrt[]{-g_{\mu\nu} J^\mu J^\nu},~ n=\frac{|J|}{\sqrt[]{-g}}\, ,$$
where $u^{\mu}_{(m)}$ is the timelike 4-velocity of matter satisfying $u^{\mu}_{(m)}u_{\mu}^{(m)}=-1$. The $\rho_{m}(n,s)$ is assumed to be prescribed as a function of $n$, the particle number density, and $s$, the entropy density per particle. The Lagrange multipliers are introduced to account for the various thermodynamic constraints imposed on the fluid. In this work, we do not require the thermodynamic aspects and, consequently, those multipliers will not be discussed in detail.
The Lagrangian for the scalar field, which we will be using for dark energy in our models, is given by,
\begin{eqnarray}\label{lagrangian-dark-energy}
\mathcal{L}_{\phi}=-\sqrt[]{-g}~\left(\frac{1}{2}\epsilon_d\partial_{\mu}\phi \partial^{\mu}\phi+V\left(\phi\right)\right)\, ,
\end{eqnarray}
where $\epsilon_d=1,-1$ are for quintessence and phantom-like scalar field, respectively and $V(\phi)$ is the potential of the scalar field $\phi$. Lastly, the Lagrangian for the interacting sector is
\begin{eqnarray}\label{lagrangian-interaction}
\mathcal{L}_{\text{int}}=-\sqrt[]{-g}f\left(n,s,\phi\right)\, ,
\end{eqnarray}
where $f(n,s,\phi)$ is an arbitrary function of $n,~s$ and $\phi$. In this work, we will consider $f(n,\phi)={\rho}_{\text{int}}$, where $\rho_{\rm{int}}$ is the energy density of the interaction term. It is assumed that there are no thermal effects in the dark sector interaction and consequently, $s$ does not play any active role henceforth. The full action for the interacting scalar field dark energy model is given by Eq.~\eqref{action}, including the gravitational, matter, scalar-field, and interaction sectors, with $\Lambda=0$. The corresponding non-interacting scalar-field model is obtained by removing the interaction sector, $\mathcal{L}_{\rm int}$, while retaining $\Lambda=0$. We use this model as a reference to isolate the effects arising from the interaction between dark matter and dark energy. For comparison with the standard cosmological scenario, we also consider the $\Lambda$CDM model, which is recovered by setting the scalar-field and interaction sectors to zero and retaining the cosmological constant $\Lambda$ in the gravitational sector.

The energy--momentum tensor corresponding to each constituent of the action is obtained by varying its Lagrangian density with respect to the metric tensor, and is given by,
\begin{eqnarray}\label{energy-momentum-formula}
T_{\mu\nu}=-\frac{2}{\sqrt{-g}}\frac{\delta \mathcal{L}}{\delta g^{\mu\nu}}.
\end{eqnarray}
Using this definition, the energy--momentum tensors for the individual constituents can be derived and are given below.
\begin{itemize}
\item \textbf{Dark matter:}
The dark matter is modeled as a perfect fluid, whose energy--momentum tensor is obtained by substituting Eq. \eqref{lagrangian-matter} into Eq. \eqref{energy-momentum-formula}, using the thermodynamic relation $P_m=n\,\partial\rho_m/\partial n-\rho_m$, and subsequently raising the first index:
\begin{equation}\label{EM-matter}
T^{\mu}{}_{\nu\left(m\right)} =
\left(\rho_{m} + P_{m}\right)u^{\mu}_{(m)}u_{\nu}^{(m)}
+ P_{m}\,\delta^{\mu}{}_{\nu},
\end{equation}
where $\rho_{m}$ and $P_{m}$ denote the dark matter energy density and pressure, respectively.
\item \textbf{Scalar-field dark energy:}
The energy-momentum tensor of the scalar field is obtained by substituting Eq. \eqref{lagrangian-dark-energy} into Eq. \eqref{energy-momentum-formula} and subsequently raising the first index, yielding the energy--momentum tensor for dark energy,
\begin{equation}\label{EM-scalarfield-DE}
T^\mu{}_{\nu(\phi)}
=
\epsilon_d\,\partial^\mu\phi\,\partial_\nu\phi
-
\delta^\mu{}_\nu
\left(
\frac{1}{2}\epsilon_d\,\partial_\alpha\phi\,\partial^\alpha\phi
+
V(\phi)
\right).
\end{equation}
For comparison, this can be equivalently written in the perfect-fluid form as,
\begin{eqnarray}\label{EM-scalarfield-DE-fluid-form}
T^{\mu}{}_{\nu(\phi)}
=
\left(\rho_{\phi}+P_{\phi}\right)
u^{\mu}_{(\phi)}u_{\nu}^{(\phi)}
+
P_{\phi}\delta^{\mu}{}_{\nu},
\end{eqnarray}
where the effective four-velocity and the corresponding energy density and pressure are defined as,
\begin{equation}
u_{\mu}^{(\phi)}
=
-\frac{\partial_{\mu}\phi}
{\sqrt{-\partial^{\alpha}\phi\partial_{\alpha}\phi}},
\quad
\rho_\phi
=
\frac{1}{2a^{2}}\epsilon_{d}{\phi'}^{2}
+
V(\phi),
\qquad
P_\phi
=
\frac{1}{2a^{2}}\epsilon_{d}{\phi'}^{2}
-
V(\phi).
\end{equation}
Throughout this work, a prime denotes the derivative with respect to the conformal time.
\item \textbf{Interaction component:}
In the present work, the interaction between dark matter and dark energy is described by an effective energy--momentum tensor assumed to take the perfect-fluid form. It is obtained by substituting Eq. \eqref{lagrangian-interaction} into Eq. \eqref{energy-momentum-formula}, using the thermodynamic relation $P_{\rm int}=n\,\partial\rho_{\rm int}/\partial n-\rho_{\rm int}$:
\begin{equation}\label{EM-interaction}
T^{\mu}{}_{\nu\left(\mathrm{int}\right)}
=
\left(\rho_{\mathrm{int}}+P_{\mathrm{int}}\right)
U^{\mu}U_{\nu}+P_{\mathrm{int}}\delta^{\mu}{}_{\nu},
\end{equation}
where $\rho_{\mathrm{int}}$ and $P_{\mathrm{int}}$ represent the effective interaction energy density and pressure, respectively, and $U_{\mu}$ is the four-velocity of the interaction component.

For the specific model considered in this work, the interaction energy density and pressure are taken as
\begin{equation}\label{Eint}
\rho_{\mathrm{int}}
=
\gamma\rho_m^{\alpha}e^{-\beta\phi},
\qquad
P_{\mathrm{int}}
=
(\alpha-1)\rho_{\mathrm{int}},
\end{equation}
where $\gamma$, $\alpha$, and $\beta$ are the model parameters that characterize the strength and form of the coupling between dark matter and the scalar field. Throughout this work, we set $\alpha=1$. This form of interaction has been previously employed in the literature in the context of interacting dark sector models \cite{Boehmer:2015kta,Boehmer:2015sha, Dutta:2017kch, Dutta:2017wfd}. The same interaction prescription was also adopted in the earlier work \cite{Saha:2024xbg}, which motivates its use in the present analysis. The reason to choose such an interaction is related to the fact that this is perhaps one of the simplest forms of dark sector coupling which yields an unambiguous  fixed point analysis for the late time universe, as discussed in the previous references. Choosing more complicated interactions \cite{Barros:2019rdv,Shahalam:2015sja,Shahalam:2017fqt} may not always produce proper late time cosmology results. Since the interaction energy density is directly proportional to the cold dark matter energy density,  we will not assume the interaction energy density to be an independent quantity, it will be absorbed in an effective matter energy density component. The effective matter energy density is therefore given by,
\begin{equation}\label{effective-rho}
  \rho_m^{\text{eff}} = \rho_{m} + \gamma \rho_{m} e^{-\beta \phi}.  
\end{equation} 
Consequently, the interaction is assumed to co-move with the cold dark matter fluid, so that both components share the same four-velocity, 
\begin{equation}\label{fourvelocity}
   U^{\mu}=u^{\mu}_{(m)}.
\end{equation}
Now, using Eqs. \eqref{EM-matter}, \eqref{EM-interaction}, \eqref{Eint}, \eqref{effective-rho}, \eqref{fourvelocity}, the total effective matter energy-momentum tensor can be written as,
\begin{equation}\label{EM-effetivematter}
T^{\mu\left(\rm{eff}\right)}{}_{\nu\left(m\right)} =T^{\mu}{}_{\nu\left(m\right)}+T^{\mu}{}_{\nu\left(\mathrm{int}\right)}=
\left(\rho_{m}^{\rm{eff}} + P_{m}^{\rm{eff}}\right)u^{\mu}_{(m)}u_{\nu}^{(m)}
+ P_{m}^{\rm{eff}}\,\delta^{\mu}{}_{\nu}.
\end{equation}
\end{itemize}
From the action in Eq. \eqref{action}, the Einstein field equations for the systems can be obtained in terms of the total energy--momentum tensor as,
\begin{equation}\label{einsteinequation}
G^{\mu}{}_{\nu}=T^{\mu}{}_{\nu(\rm{tot})},
\end{equation}
where $G^{\mu}{}_{\nu}$ is the Einstein tensor and $T^{\mu}{}_{\nu(\rm{tot})}$ is the total energy--momentum tensor, which includes contributions from dark matter, dark energy, and the interaction component, when present.
In addition to the Einstein equation, the conservation equation will also be needed for the solution of the dynamical equations of the systems. For the interacting scalar-field model, the total energy--momentum tensor is conserved and satisfies,
\begin{equation}\label{conservation-interacting}
\nabla_{\mu}T^{\mu}{}_{\nu\left(\text{tot}\right)}
=
\nabla_{\mu}T^{\mu\left(\rm{eff}\right)}{}_{\nu\left(m\right)}
+
\nabla_{\mu}T^{\mu}{}_{\nu\left(\phi\right)}
=0 ,
\end{equation}
The individual sectors need not be conserved separately because the interaction allows for an exchange of energy and momentum between the dark matter and scalar-field sectors. The non-interacting scalar-field model is obtained by removing the interaction contribution from the action. In this case, the dark matter and scalar-field sectors evolve independently and are separately conserved,
\begin{equation}\label{conservation-noninteracting}
\nabla_{\mu} T^{\mu}{}_{\nu\left(m\right)} = 0, 
\qquad
\nabla_{\mu} T^{\mu}{}_{\nu\left(\phi\right)} = 0.
\end{equation}
Finally, the $\Lambda$CDM model is obtained in the absence of the scalar-field and interaction sectors, with the dark matter energy--momentum tensor satisfying
\begin{equation}\label{conservation-lambdacdm}
\nabla_{\mu} T^{\mu}{}_{\nu\left(m\right)} = 0.
\end{equation}
\subsection{Perturbation variables}
\label{pvaiables}

Including scalar perturbations  and working in the Newtonian gauge, the perturbed metric takes the form \cite{Bertschinger:2001is,lesgourgues2013},
\begin{equation}
\label{perturbedmetric}
ds^{2} = a^{2}\left(\tau\right)\Big[
-\left(1 + 2\Psi\left(\boldsymbol{x},\tau\right)\right)d\tau^{2}+ \left(1 - 2\Phi\left(\boldsymbol{x},\tau\right)\right)
\delta_{ij}\,dx^{i}dx^{j}
\Big],
\end{equation}
where $\Phi\left(\boldsymbol{x},\tau\right)$ and $\Psi\left(\boldsymbol{x},\tau\right)$ are the scalar metric potentials. Anisotropic stress is absent for CDM, $\Lambda$ or scalar field dark energy; also, there are no radiative components like neutrinos; therefore, the two scalar metric potentials in the Newtonian gauge coincide,
\begin{equation}\label{equalpotential}
\Psi = \Phi .
\end{equation}
The perturbed four-velocity is,
\begin{equation}\label{perturbvelocity}
u^{\mu} = \frac{1}{a}\left(1 - \Phi,\, v^{i}\right),
\end{equation}
where $v^{i}$ denotes the $i$-th spatial component of the peculiar
velocity of the fluid, with $i=1,2,3$. At the background level, the
fluid is comoving with the cosmological expansion, and hence
$\bar{u}^{\mu}=(1/a,0,0,0)$. The perturbed
four-velocity satisfies the normalization condition
$u^{\mu}u_{\mu}=-1$ to linear order. This follows for all the components: CDM, scalar field dark energy or the interaction term's four-velocity. Using the perturbed metric given in Eq. \eqref{perturbedmetric} and the form of the perturbed four-velocity from Eq. \eqref{perturbvelocity} in the energy--momentum tensors defined in Eqs. \eqref{EM-matter}, \eqref{EM-scalarfield-DE-fluid-form}, and \eqref{EM-interaction}, and retaining terms up to first order in the perturbations, the non-vanishing components of the energy--momentum tensor for each component $A=\{m,\phi,\mathrm{int}\}$ are given by,
\begin{align}\label{perturbform}
T^{0}{}_{0(A)} &= -\left(\bar{\rho}_{A}+\delta\rho_{A}\right), \nonumber\\
T^{0}{}_{i(A)} &= \left(\bar{\rho}_{A}+\bar{P}_{A}\right)v_{i(A)}, \nonumber\\
T^{i}{}_{j(A)} &= \left(\bar{P}_{A}+\delta P_{A}\right)\delta^{i}{}_{j}.
\end{align}
Here, $\bar{\rho}_{A}$ and $\bar{P}_{A}$ denote the background energy
density and pressure, while $\delta\rho_{A}$ and $\delta P_{A}$ are
their respective first-order perturbations. The quantity
$v_{i(A)}$ denotes the peculiar velocity perturbation of component $A$.

CDM is modeled as a pressureless perfect fluid, which makes the perturbation pressure $\delta P_m= 0$.  At the linear perturbation level, the perturbations of CDM are completely characterized by the density contrast and the velocity perturbation. The density contrast is defined as the fractional perturbation in the cold dark matter energy density,
\begin{equation}\label{densitycontrastdm}
\delta_m = \frac{\delta\rho_m}{\bar{\rho}_m},
\end{equation}
where $\delta\rho_m$ denotes the perturbation in the cold dark matter energy density. The velocity perturbation is described by the divergence of the peculiar velocity field $\mathbf{v}_m$ of the cold dark matter fluid,
\begin{equation}\label{thetadm}
\theta_m \equiv \nabla \cdot \mathbf{v}_m.
\end{equation}
In scalar-field dark energy models, dark energy can also develop spatial perturbations and therefore contribute dynamically to the evolution of cosmic structures. The scalar field is decomposed into a background and a perturbation as,
\begin{equation}\label{perturbation}
\phi\left(\boldsymbol{x},t\right)
= \bar{\phi}\left(t\right) + \delta\phi\left(\boldsymbol{x},t\right),
\end{equation}
The linear perturbations in the scalar-field energy density and pressure, together with the $0i$ component of the scalar-field energy--momentum tensor, are obtained by substituting the perturbed scalar field from Eq. \eqref{perturbation} and the perturbed metric components from Eq. \eqref{perturbedmetric} into Eq. \eqref{EM-scalarfield-DE}, retaining terms up to first order in the perturbations and isolating the perturbative contributions to the $0$-$0$, $i$-$j$, and $0$-$i$ components. The expressions are,
\begin{equation}\label{delta_rho_phi}
\delta\rho_{\phi}
=
\frac{\epsilon_{d}}{a^{2}}
\left(
\bar{\phi}'\,\delta\phi'
-
\bar{\phi}'^{\,2}\Phi
\right)
+
\frac{\mathrm{d}V}{\mathrm{d}\bar{\phi}}\delta\phi,
\end{equation}
\begin{equation}\label{delta_P_phi}
\delta P_{\phi}
=
\frac{\epsilon_{d}}{a^{2}}
\left(
\bar{\phi}'\,\delta\phi'
-
\bar{\phi}'^{\,2}\Phi
\right)
-
\frac{\mathrm{d}V}{\mathrm{d}\bar{\phi}}\delta\phi,
\end{equation}
and
\begin{equation}\label{T0i_phi}
T^{0}{}_{i(\phi)}
=
-\frac{\epsilon_{d}\bar{\phi}'}{a^{2}}\partial_{i}\delta\phi.
\end{equation}
The scalar field density contrast is defined by,
\begin{equation}\label{densitycontrastphi}
\delta_{\phi} = \frac{\delta\rho_{\phi}}{\bar{\rho}_{\phi}}.
\end{equation}
Using Eq. \eqref{T0i_phi} together with the perturbed $T^{0}{}_{i}$ component given in Eq. \eqref{perturbform}, and defining the velocity divergence as $\theta_\phi \equiv \partial^{i}v_{i_\phi}$, where $v_{i_\phi}$ denotes the $i$th component of the peculiar velocity of the scalar-field dark energy, the corresponding relation in real space is obtained. Transforming this relation to Fourier space and expressing the conformal-time derivative of $\phi$ in terms of the derivative with respect to $\ln a$, the expression becomes,
\begin{equation}\label{theta_phi}
\theta_\phi =\frac{\epsilon_{d}k^{2}\mathcal{H} \delta \phi}{a^{2}\left(\bar{\rho}_{\phi}+\bar{P}_{\phi}\right)}\frac{\mathrm{d}\bar{\phi}}{\mathrm{d}\ln a}.
\end{equation}
Since the interaction contributes to the effective matter component, we substitute $\rho_m=\bar{\rho}_m+\delta\rho_m$ and Eq. \eqref{perturbation} into Eq. \eqref{effective-rho}. Expanding the resulting expression in the perturbations and retaining terms up to linear order, we isolate the perturbative contribution by subtracting the background part and substituted Eq. \eqref{Eint}. The resulting effective matter density perturbation is then given by,
\begin{equation}\label{deltarhoeff}
\delta\rho_m^{\mathrm{eff}}
=
\bar{\rho}^{\rm{eff}}_{m}\delta_m
-\beta \bar{\rho}_{\rm{int}} \delta\phi.
\end{equation}
For $\alpha=1$, the interaction pressure in Eq. \eqref{Eint} vanishes, $P_{\rm {int}}=0$. Since the cold dark matter is also pressureless, $P_m=0$, the effective matter pressure and its perturbation consequently vanish, i.e., $\delta P^{\rm eff}=0$. Furthermore, since the interaction and cold dark matter share the same four-velocity as mentioned in Eq. \eqref{fourvelocity}, the velocity divergence of the interaction component satisfies
\begin{equation}\label{velocityequality}
\theta_{\mathrm{int}}=\theta_m.
\end{equation}
The effective matter density contrast is defined by,
\begin{align}\label{densitycontrasteff}
\delta_{m}^{\rm{eff}}
&=\frac{\delta \rho_{m}^{\rm{eff}}}{\bar{\rho}_{m}^{\rm{eff}}}\\ \label{densitycontrasteff2}
&=
\delta_{m}
-\mathcal{C}\delta\phi
\end{align}
where the second equality comes from using Eqs. \eqref{deltarhoeff} and \eqref{effective-rho} and $\mathcal{C}=\frac{\beta\bar{\rho}_{\rm{int}}}
{\bar{\rho}^{\rm{eff}}_{m}}$.

All the dynamics studied in this work were derived from the same underlying metric and energy--momentum tensors, as defined above. The full set of linearized Einstein and conservation equations follows directly from this framework, whereas the sub-horizon (SHA) and quasi-static (QSA) equations are obtained by applying controlled approximations in the perturbation part of the system. No additional assumptions are introduced at the level of the metric, matter content, or gauge choice; the differences between the full, SHA, and QSA treatments arise solely from the effect of the different dynamical approximations imposed on linear perturbation equations.
\section{The approximations}\label{Approximations}

The sub-horizon limit corresponds to considering perturbation modes whose wavelengths are significantly smaller than the Hubble radius. In conformal time \(\tau\), this corresponds to modes with comoving wavenumber \(k\) satisfying
\begin{equation}\label{SHAcondition}
    k \gg \mathcal{H}, \quad \rm{or} \quad \epsilon \equiv \frac{\mathcal{H}}{k} \ll 1,
\end{equation}
where \(\mathcal H\equiv a'/a\) is the conformal Hubble parameter. The parameter \(\epsilon\) characterizes the separation between the Hubble and perturbation length scales. The condition \(k\gg\mathcal H\), by itself, represents only a separation of spatial scales and does not imply a particular temporal behaviour of the perturbations. If the Hubble time \(\mathcal H^{-1}\) is the only relevant characteristic time scale governing their evolution, temporal derivatives are expected to scale as \(\partial_\tau\sim\mathcal H\), whereas spatial derivatives scale as \(\nabla\sim k\). The SHA may then be implemented by neglecting terms containing temporal derivatives when they are suppressed by powers of \(\mathcal H/k\) relative to the corresponding spatial-gradient terms. This does not require the gravitational potentials themselves to be slowly varying in the quasi-static sense: one may have \(\Phi'/(\mathcal H\Phi)\sim O(1)\), while derivative contributions involving \(\Phi'\) remain suppressed relative to \(k\)-enhanced spatial-gradient terms. In cosmological models containing additional dynamical degrees of freedom, however, additional characteristic time scales may arise, so that the ordering \(\partial_\tau\sim\mathcal H\) is not guaranteed. The accuracy of the SHA must therefore be established rather than assumed in such models.
%

The quasi-static approximation (QSA) introduces an additional dynamical assumption concerning the evolution of the metric perturbations. Since \(\Psi=\Phi\) according to Eq. (1.22), it is sufficient to consider a single gravitational potential. The QSA applies when the characteristic evolution timescale of \(\Phi\) is much longer than the Hubble timescale \(\mathcal H^{-1}\), so that the potential may be regarded as approximately time-independent over cosmological expansion timescales. In conformal time, this condition may be quantified through
\begin{equation}\label{QSAcondition}
    \epsilon_\Phi \equiv \frac{\Phi'}{\mathcal{H}\Phi}= \frac{1}{\Phi}\frac{\mathrm{d}\Phi}{\mathrm{d}\ln a},
\end{equation}
with
$$ |\epsilon_\Phi|\ll1. $$
Under this condition, time-derivative terms involving the metric potential may be neglected in the linearized Einstein equations. In the present work, the QSA is imposed only on the metric perturbations; the time evolution of the scalar-field perturbation is retained.

In the standard $\Lambda$CDM model, the SHA and the QSA are typically valid during the matter-dominated era on scales well within the horizon, where the gravitational potential varies slowly with time. Under these conditions, the Einstein equations reduce to their Newtonian limit, yielding the Poisson equation along with the continuity and Euler equations that govern matter perturbations. We apply the same approximations to the interacting scalar-field dark energy model to investigate how the dark-sector interaction modifies this standard Newtonian behavior. For comparison, we also consider the corresponding non-interacting scalar-field model to isolate the effects arising from the scalar-field dynamics itself.

Other than these approximation parameters, another parameter used throughout this work is $\epsilon_{d}$. This parameter should not be confused with the two approximation parameters. It is introduced to distinguish between quintessence and phantom-like scalar field dark energy, taking the constant value $\epsilon_{d}=1,-1$ for quintessence and phantom fields, respectively.
\section{Background cosmology and the parameter set}
\label{bcp}

In this section we specify the various background cosmological scenarios we use in this paper. Later the cosmological perturbations over these backgrounds will be presented.
\subsection{Interacting scalar field dark Energy}\label{InteractingScalarFieldDE}

In the interacting scalar field dark energy model, the cosmic fluid consists of CDM, scalar field dark energy, and the interaction term. Substituting Eqs. \eqref{EM-scalarfield-DE-fluid-form}, and \eqref{EM-effetivematter}, into the Einstein equations given by Eq. \eqref{einsteinequation}, setting $\Lambda=0$, and considering all the $00$ components, the Friedmann equation governing the background expansion of the Universe can be obtained:
\begin{equation}\label{Friedmann-interacting}
\mathcal{H}^{2}=\frac{a^{2}}{3}
\left( \bar{\rho}^{\rm{eff}}_m + \bar{\rho}_\phi \right)\,.
\end{equation}
Again, substituting the energy-momentum tensor equations, Eqs. \eqref{EM-scalarfield-DE} and \eqref{EM-effetivematter} into the total energy conservation equation, Eq. \eqref{conservation-interacting}, and separating the contributions from effective matter and the scalar field, the following background evolution equations are obtained. For pressureless effective dark matter, the conservation equation is
\begin{equation}\label{continuitybackgroudinteracting}
\bar{\rho}^{\rm{eff}'}_{m} =- 3\mathcal{H}\bar{\rho}^{\rm{eff}}_{m}+ \frac{\partial \bar{\rho}_{int}}{\partial \bar{\phi}}\bar{\phi}',
\end{equation}
and the Klein-Gordon equation, after expressing the conformal-time derivatives in terms of derivatives with respect to $\ln a$, can be written as,
\begin{equation}\label{KGbackgroundinteracting}
\frac{\mathrm{d}^{2}\bar{\phi}}{\mathrm{d}\left(\ln a\right)^{2}}
=- \left(2 + \frac{\mathrm{d}\ln \mathcal{H}}{\mathrm{d}\ln a}\right)
\frac{\mathrm{d}\bar{\phi}}{\mathrm{d}\ln a}-\frac{\epsilon_{d}a^{2}}{\mathcal{H}^{2}}
\left(\frac{\partial V}{\partial \bar{\phi}}+ \frac{\partial \bar{\rho}_{int}}{\partial \bar{\phi}}\right).
\end{equation}
Starting from the Friedmann equation, Eq.~\eqref{Friedmann-interacting}, differentiating it with respect to $\ln a$ and using the total background energy conservation equation, Eq.~\eqref{conservation-interacting}, for algebraic simplification, the expression becomes,
\begin{equation}\label{logfriedmanneq}
\frac{\mathrm{d}\ln \mathcal{H}}{\mathrm{d}\ln a}
= -\frac{1}{2}\left(1 + 3\omega_{\mathrm{tot}}\right),
\end{equation}
with $\omega_{\mathrm{tot}}=\bar{P}_{\phi}/(\bar{\rho}^{\rm{eff}}_{m}+\bar{\rho}_{\phi})$ denoting the effective equation-of-state parameter of the total cosmic fluid.
The time-dependent density parameters are given by,
\begin{equation}\label{timedensityparameters}
\Omega_m\left(a\right) =
\frac{a^{2}\bar{\rho}^{\rm{eff}}_m(a)}
{3\mathcal{H}^{2}},
\quad \quad \Omega_\phi(a)
=
\frac{a^{2}\bar{\rho}_\phi(a)}
{{3\mathcal{H}^{2}}}.
\end{equation}
At the present epoch ($a=1$), the density parameters satisfy
\begin{equation}
\Omega_{m_{0}}+\Omega_{\phi_{0}}=1,
\qquad
\Omega_{\phi_{0}}=\Omega_{\Lambda_{0}},
\end{equation}
We adopt the present-day density parameters $\Omega_{m_{0}}=0.3$ and $\Omega_{\Lambda_{0}}=0.7$, consistent with the \textit{Planck} 2018 results \cite{Planck:2018vyg}, which have been kept the same throughout all the models in our work. Although the models considered in this work explicitly include only the cold dark matter component, we neglect the distinction between baryonic matter and cold dark matter. Consequently, the quoted value of $\Omega_{m_{0}}$, which corresponds to the total matter density parameter in the \textit{Planck} analysis, is taken to represent the effective cold dark matter density in our work.
\begin{figure}[htbp]
        \centering
        \includegraphics[width=0.45\linewidth]{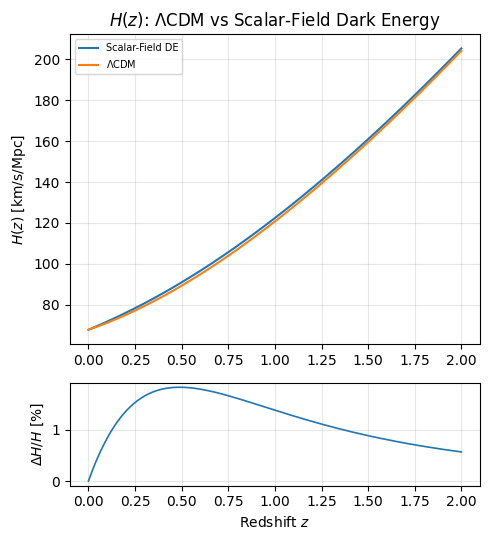}
        \caption{Evolution of the Hubble parameter $H(z)$ for the scalar-field dark energy model and $\Lambda$CDM (top), and their relative percentage difference (bottom).}
        \label{H_match_LCDM}
\end{figure}
\subsection{Non-interacting scalar field dark energy}\label{MinimalSFDE}

In the non-interacting scalar-field dark energy model, the intrinsic properties of the cold dark matter (CDM) and scalar-field dark energy components remain unchanged, while the additional contributions arising from the interaction term are absent. Consequently, all the background equations of the interacting model reduce to those of the non-interacting case by setting the interaction parameter $\gamma=0$. Equivalently, the equations governing the CDM and scalar-field dark energy components can be obtained independently without including the interaction sector in the action. 

Substituting Eqs. \eqref{EM-matter} and \eqref{EM-scalarfield-DE-fluid-form} into the Einstein equations given by Eq. \eqref{einsteinequation}, setting $\Lambda=0$, and considering the $00$ component, the Friedmann equation governing the background expansion of the Universe for this model can be obtained.
\begin{equation}
\mathcal{H}^{2}
= \frac{a^{2}}{3}
\left( \bar{\rho}_m + \bar{\rho}_\phi \right),
\end{equation}
The background energy conservation equations follow from the separate conservation of matter and scalar-field energy--momentum tensors. Substituting Eqs. \eqref{EM-matter} and \eqref{EM-scalarfield-DE} into the separate energy conservation equations, in Eq. \eqref{conservation-noninteracting}. The two conservation equations become,
\begin{align}
  \bar{\rho}'_{m} &=- 3\mathcal{H}\bar{\rho}_{m},  \\[6pt]
\frac{\mathrm{d}^{2}\bar{\phi}}{\mathrm{d}\left(\ln a\right)^{2}}
&=- \left(2 + \frac{\mathrm{d}\ln \mathcal{H}}{\mathrm{d}\ln a}\right)
 \frac{\mathrm{d}\bar{\phi}}{\mathrm{d}\ln a}
- \frac{\epsilon_{d}a^{2}}{\mathcal{H}^{2}}
\frac{\mathrm{d}V}{\mathrm{d}\bar{\phi}}.
\end{align}
where the expression for $\mathrm{d}\ln \mathcal{H}/\mathrm{d}\ln a$ remains the same as in Eq.~(\ref{logfriedmanneq}), with $\omega_{\mathrm{tot}}=\bar{P}_{\phi}/(\bar{\rho}_{m}+\bar{\rho}_{\phi})$ denoting the effective equation-of-state parameter of the total cosmic fluid. The dark-energy density parameter remains unchanged and is given by
Eq.~\eqref{timedensityparameters}, while the matter density parameter is
modified to,
\begin{equation}\label{matterdensityparameter}
\Omega_m(a)
=
\frac{a^{2}\bar{\rho}_m(a)}
{3\mathcal{H}^{2}}.
\end{equation}.
\subsection{$\Lambda$CDM}\label{LambdaCDM}

In the $\Lambda$CDM model, dark energy is represented by the cosmological constant $\Lambda$, and there is no scalar-field dark energy component. Substituting the energy--momentum tensor of cold dark matter, Eq. \eqref{EM-matter}, into the Einstein equations given by Eq. \eqref{einsteinequation}, and considering the $00$ component, the Friedmann equation governing the background expansion of the Universe can be obtained as,
\begin{equation}\label{BackgroundlambdaCDM}
\mathcal{H}^{2}
= \frac{a^{2}}{3}
\left(
\bar{\rho}_m + \Lambda
\right).
\end{equation}
The background evolution of the CDM density is obtained by substituting Eq. \eqref{EM-matter} into Eq. \eqref{conservation-lambdacdm}.
\begin{equation}
\bar{\rho}'_{m} =- 3\mathcal{H}\bar{\rho}_{m}
\end{equation}
and the matter density parameter remains unchanged from the non-interacting model and is given by
Eq.~\eqref{matterdensityparameter}.
\section{Perturbation Equations}
\label{perturb}

In this work, we derive the linear perturbation equations for the interacting scalar-field dark energy model considered here. Since the corresponding perturbation equations for the $\Lambda$CDM and non-interacting scalar-field models are already well established in the literature \cite{Unnikrishnan:2008qe, Mukhanov:1990me, Hwang:2001fb, Hwang:2018vpz}, we do not re-derive them. Instead, these known results are used as reference cases for comparison with the interacting model.

At the linear level, Einstein's equations provide three independent sets of equations corresponding to the $00$, $0i$, and $ij$ components. However, the $ij$ component can be obtained from the $00$ and $0i$ components together with the energy--momentum conservation equations, and therefore does not provide any additional independent information. Consequently, in all the models, we consider only the $00$ and $0i$ components of the linearized Einstein equations. In addition to Einstein's equations, the conservation of the cold dark matter energy--momentum tensor,$\nabla_{\mu}\delta T^{\mu}{}_{\nu\left(m\right)}=0$, yields two equations: the continuity equation from the $\nu=0$ component and the Euler equation from the $\nu=i$ components. But, for the scalar field, the conservation equation $\nabla_{\mu}\delta T^{\mu}{}_{\nu\left(\phi\right)}=0,$  gives only one independent equation from the $\nu=0$ component and is equivalent to the perturbed Klein--Gordon equation and provides the evolution equation for the scalar-field perturbation. For the interacting scalar-field model, the individual dark matter and scalar-field energy--momentum tensors are not separately conserved. Instead, the total energy--momentum tensor satisfies
\begin{equation}\label{Eq:totalconservationInteracting}
\nabla_{\mu}
\left(
\delta T^{\mu\left(\rm{eff}\right)}{}_{\nu\left(m\right)}
+
\delta T^{\mu}{}_{\nu\left(\phi\right)}
\right)=0.
\end{equation}
We derive the corresponding continuity and Euler equations for the effective matter sector, together with the perturbed Klein--Gordon equation, from Eq. \eqref{Eq:totalconservationInteracting}. Together, these constitute the complete set of linear perturbation equations used in this work. The perturbation equations are obtained by retaining only terms up to first order in the perturbation variables. As a result, the evolution equations are linear in the perturbations, and different Fourier modes evolve independently without mode coupling. Therefore, in the equations below, we work in Fourier space, where $k$ denotes the comoving wavenumber, and we also express all derivatives with respect to $\ln a$, as it is convenient for implementing the sub-horizon and quasi-static approximations.

In the following subsections, we first present the complete set of linearized perturbation equations and subsequently apply the sub-horizon and quasi-static approximations to derive the corresponding evolution equations. Throughout the analysis, the approximations are implemented by taking the strict limits $\epsilon \ll 1$ and $\epsilon_{\Phi} \ll 1$ using Eqs. (\ref{SHAcondition}) and (\ref{QSAcondition}), such that we evaluate the system to leading order, keeping only the zeroth-order terms ($\mathcal{O}(\epsilon^0)$ and $\mathcal{O}(\epsilon_{\Phi}^0)$) and consistently neglecting all higher-order terms in the sub-horizon and quasi-static approximations, respectively.
\begin{figure*}
\subfigure[Redshift evolution of $D$ for $k=0.01Mpc^{-1}$]
{\includegraphics[width=0.47\textwidth,height=52mm]{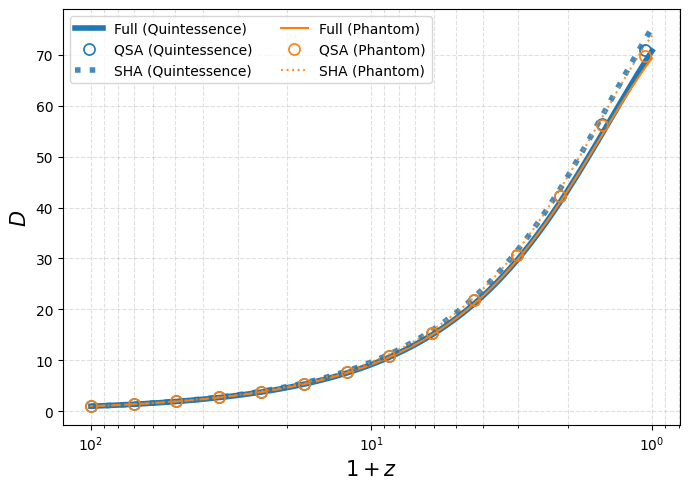}\label{Nonminimaldeltak_0.01}}
\hspace{0.2cm}
\subfigure[Redshift evolution of $f$ for $k=0.01Mpc^{-1}$]
{\includegraphics[width=0.47\textwidth,height=52mm]{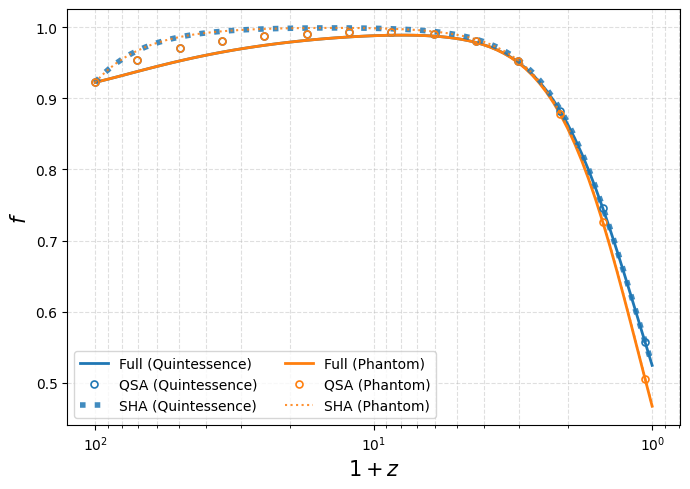}\label{Nonminimal_f_k_0.01}}

\subfigure[Redshift evolution of $D$ for $k=0.005Mpc^{-1}$]
{\includegraphics[width=0.47\textwidth,height=52mm]{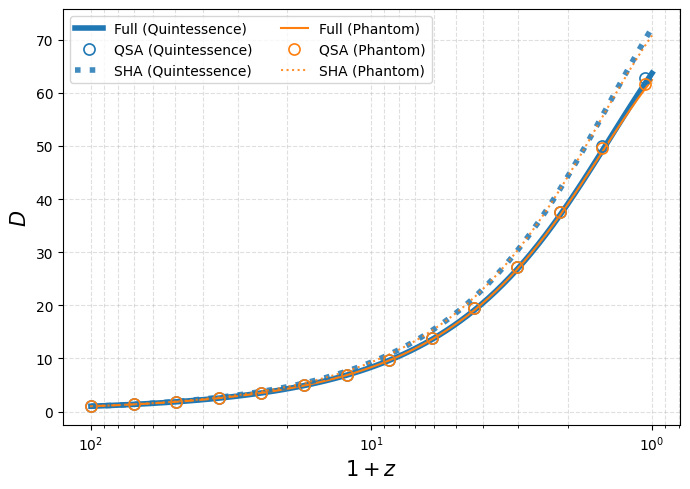}\label{Nonminimaldeltak_0.005}}
\hspace{0.2cm}
\subfigure[Redshift evolution of $f$ for $k=0.005Mpc^{-1}$ ]
{\includegraphics[width=0.47\textwidth,height=52mm]{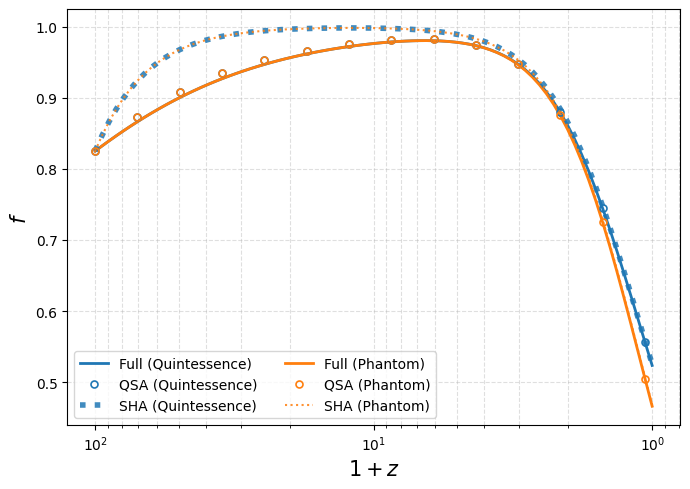}\label{Nonminimal_f_k_0.005}}

\subfigure[Redshift evolution of $D$ for $k=0.003Mpc^{-1}$]
{\includegraphics[width=0.47\textwidth,height=52mm]{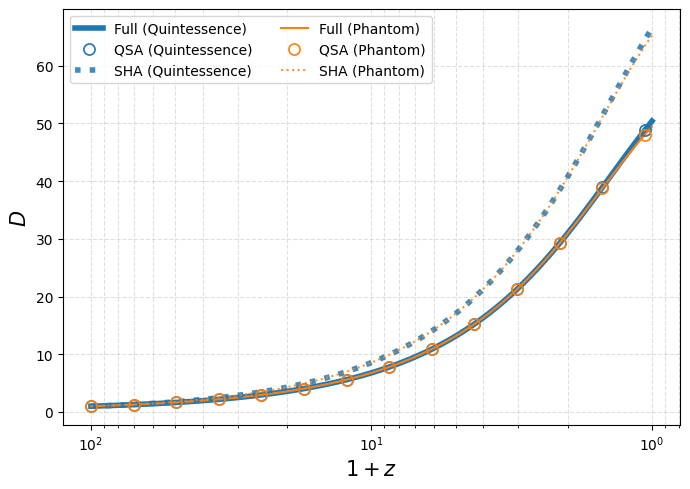}\label{Nonminimaldeltak_0.003}}
\hspace{0.2cm}
\subfigure[Redshift evolution of $f$ for $k=0.003Mpc^{-1}$]
{\includegraphics[width=0.47\textwidth,height=52mm]{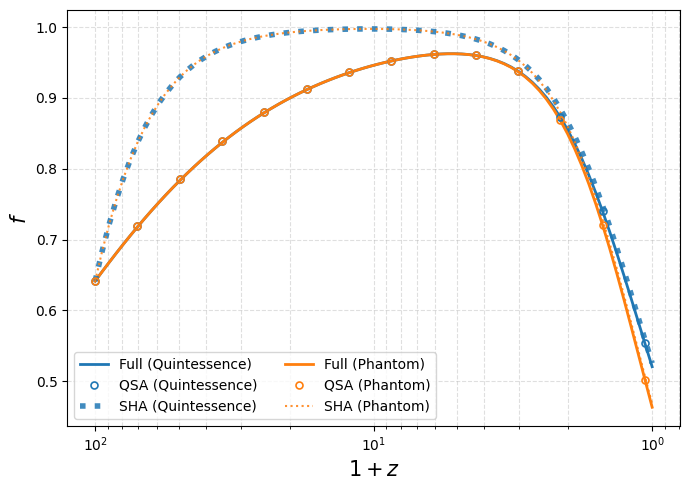}\label{Nonminimal_f_k_0.003}}

\caption{Redshift evolution of the cold dark matter linear growth factor,
$D(a,k)=\delta_m(a,k)/\delta_m(a_{\rm ini},k)$ (left panels), and
the corresponding linear growth rate,
$f(a,k)=\mathrm{d}\ln D/\mathrm{d}\ln a$ (right panels), for an
interacting scalar-field dark energy model with quintessence and phantom
fields. Results are shown for three representative comoving wavenumbers,
$k=0.01$, $0.005$, and $0.003\,\mathrm{Mpc}^{-1}$, with interaction
parameters $\beta=0.001$ and $\gamma=0.04$. For each wavenumber, the
full linear solution and the sub-horizon approximation
(SHA) and the quasi-static approximation (QSA) are plotted. As the $k$ value decreases, the SHA solution deviates more from the full solution.}
\label{Nonminimal_k}
\end{figure*}
\subsection{Interacting scalar field dark energy}

In this subsection we discuss the various aspects of the interacting scalar dark energy model used in our paper.
\subsubsection{Full linearized perturbation equations}

First we derive the full set of linear perturbation equations for the interacting scalar-field dark energy model, which consists of Einstein equations and conservation equations.
\subsubsection*{Einstein equations}
Using the perturbed metric in Eq.~\eqref{perturbedmetric} and the energy-momentum tensors in Eq.~\eqref{EM-effetivematter}, and \eqref{EM-scalarfield-DE-fluid-form}, the $00$ and $0i$ components of the Einstein equations in Eq.~\eqref{einsteinequation} can be derived.
The $00$ component of the linearized Einstein equations is given by,
\begin{align}
-k^2 \Phi
-3\mathcal{H}^2
\left(
\frac{\mathrm{d}\Phi}{\mathrm{d}\ln a}+\Phi
\right)
&=
\frac{a^2}{2}
\left(
\delta\rho_m^{\mathrm{eff}}+\delta\rho_\phi
\right)
\nonumber\\[1ex]
&=
\frac{3\mathcal{H}^2}{2}
\left[
\Omega_m\left(\delta_m
-\mathcal{C}\delta \phi\right)
+\Omega_\phi\delta_{\phi}
\right].\label{zerozeroeqlnafullSFC}
\end{align}
where, in the second line, we have expressed the two density perturbations in terms of the corresponding density contrasts using Eqs. \eqref{densitycontrasteff} and \eqref{densitycontrastphi}, then substituted the corresponding background density parameters from Eq. \eqref{timedensityparameters}, and subsequently substituted the expression for the effective matter density contrast given in Eq. \eqref{densitycontrasteff2}.

The $0i$ component of the linearized Einstein equations, after substituting the background density parameters from Eq.~\eqref{timedensityparameters} and rearranging the resulting expression, can be written as,
\begin{equation}\label{zeroieqlnafullSFC}
\frac{\mathrm{d}\Phi}{\mathrm{d}\ln a}+\Phi
=
\frac{3\mathcal{H}}{2k^2}
\Big[
\Omega_m\theta_m
+
(1+w_\phi)\Omega_\phi\theta_\phi
\Big].
\end{equation}
\subsubsection*{Energy-momentum conservation}

To obtain the conservation of the total energy--momentum tensor in the interacting scalar-field dark energy model from Eq. \eqref{Eq:totalconservationInteracting}, we evaluate the scalar-field and effective matter contributions separately and subsequently combine them. For the scalar-field sector, substituting Eq. \eqref{EM-scalarfield-DE} into the conservation equation and considering the $\nu=0$ component, retaining only the first-order perturbative contribution, followed by using the background Klein--Gordon equation for algebraic simplification, gives
\begin{align}\label{conservationkg}
\nabla_\mu \delta T^{\mu}{}_{0\left(\phi\right)}
=
{}&
- \frac{\epsilon_d\bar{\phi}'}{a^2}
\Bigg[
\delta\phi''
+2\mathcal{H}\delta\phi'
+
\left(
k^2
+\epsilon_d a^2
\frac{\mathrm{d}^2V}{\mathrm{d}\bar{\phi}^2}
\right)\delta\phi
-4\bar{\phi}'\Phi'
+2\epsilon_d a^2
\left(
\frac{\mathrm{d}V}{\mathrm{d}\bar{\phi}}
-\beta\bar{\rho}_{\mathrm{int}}
\right)\Phi
\Bigg]
\nonumber\\
&\qquad\qquad
-\beta\bar{\rho}_{\mathrm{int}}\delta\phi'.
\end{align}
Similarly, for the effective matter sector, substituting Eq. \eqref{EM-effetivematter} into the conservation equation, considering the $\nu=0$ component, and retaining only the first-order perturbative contributions yields,
\begin{align}\label{conservationeffmatter2}
\nabla_\mu \delta T^{\mu\left(\mathrm{eff}\right)}{}_{0\left(m\right)}
&=
-\delta\rho_m^{\mathrm{eff}\prime}
-3\mathcal{H}\delta\rho_m^{\mathrm{eff}}
+3\bar{\rho}_m^{\mathrm{eff}}\Phi'
-\bar{\rho}_m^{\mathrm{eff}}\theta_m.\nonumber\\
&=-\bar{\rho}_m^{\mathrm{eff}}
\left(
\delta_{m}'-3\Phi'+\theta_{m}
\right)+
\beta\bar{\rho}_{\mathrm{int}}\delta\phi'
+
\beta\bar{\phi}'\bar{\rho}_{\mathrm{int}}
\left(
\delta_{m}-\beta\delta\phi
\right).
\end{align}
The second equality comes from using the expression for $\delta\rho_m^{\mathrm{eff}}$ given in Eq. \eqref{deltarhoeff} and substituting Eqs. \eqref{Eint}, \eqref{effective-rho}.
Similarly, setting $\nu=i$ in Eq.~\eqref{EM-effetivematter}, we obtain $\nu=i$ component of the effective matter,
\begin{align}
\nabla_\mu \delta T^{\mu\left(\mathrm{eff}\right)}{}_{i\left(m\right)}
&=
\left(
\bar\rho_m^{\mathrm{eff}}v_{mi}
\right)'
+
4\mathcal H\bar\rho_m^{\mathrm{eff}}v_{mi}
+
\bar\rho_m^{\mathrm{eff}}\partial_i\Phi\,,
\nonumber \\
&=
\bar\rho_m^{\mathrm{eff}}
\left[
v_{mi}'
+
\left(
\mathcal H-\mathcal{C}\bar\phi'
\right)v_{mi}
+
\partial_i\Phi
\right]\,.
\label{effective-matter-momentum-divergence-compact}
\end{align}
Here, the second equality is obtained by expanding the derivatives and using, $\bar\rho_m^{\mathrm{eff}\,\prime}
+
4\mathcal H\bar\rho_m^{\mathrm{eff}}
=
\mathcal H\bar\rho_m^{\mathrm{eff}}
-
\beta\bar\phi'\bar\rho_{\mathrm{int}}$, which follows from the background continuity equation, Eq.~\eqref{continuitybackgroudinteracting} and using $\mathcal{C}=\frac{\beta\bar{\rho}_{\rm{int}}}
{\bar{\rho}^{\rm{eff}}_{m}}$.
Using the scalar field energy momentum tensor Eq. \eqref{EM-scalarfield-DE}, setting $\nu=i$ gives the contribution from the scalar field as,
\begin{align}\label{Eq:scalarfieldnu=i}
\nabla_\mu \delta T^{\mu}{}_{i\left(\phi\right)}
={}&
\left(
-\frac{\epsilon_d}{a^2}
\bar\phi'\partial_i\delta\phi
\right)'
-
\frac{4\mathcal H\epsilon_d}{a^2}
\bar\phi'\partial_i\delta\phi+
\partial_i\delta P_\phi
+
\frac{\epsilon_d\bar\phi'^2}{a^2}
\partial_i\Phi.
\end{align}
The first term on the right hand side can be written as:
\begin{align}
\left(
-\frac{\epsilon_d}{a^2}
\bar\phi'\partial_i\delta\phi
\right)'
={}&
-\frac{\epsilon_d}{a^2}
\Big[
\bar\phi''\partial_i\delta\phi
+
\bar\phi'\partial_i\delta\phi'
-2\mathcal H\bar\phi'
\partial_i\delta\phi
\Big].
\end{align}
The pressure-gradient term using Eq. \eqref{delta_P_phi} becomes,
\begin{align}
\partial_i\delta P_\phi
={}&
\frac{\epsilon_d}{a^2}
\left[
\bar\phi'\partial_i\delta\phi'
-
\bar\phi'^2\partial_i\Phi
\right]-
\frac{\mathrm{d}V}{\mathrm{d}\bar\phi}
\partial_i\delta\phi.
\end{align}
After substituting these expressions, the terms proportional to \(\partial_i\delta\phi'\) and \(\partial_i\Phi\) cancel. Subsequently, using the background Klein–Gordon equation, Eq.~\eqref{KGbackgroundinteracting}, Eq.~\eqref{Eq:scalarfieldnu=i} reduces to,
\begin{equation}
\nabla_\mu \delta T^{\mu}{}_{i\left(\phi\right)}
=-\beta\bar\rho_{\mathrm{int}}
\partial_i\delta\phi.
\label{scalar-momentum-divergence}
\end{equation}
Now, using Eqs. \eqref{effective-matter-momentum-divergence-compact}, and \eqref{scalar-momentum-divergence}, in Eq. \eqref{Eq:totalconservationInteracting}, we obtained, 
\begin{equation}\label{velocityeqinteracting}
\bar{\rho}^{\rm{eff}}_{m}\Big[v_{mi}'
+
\left(
\mathcal H-\mathcal{C}\bar\phi'
\right)v_{mi}
+
\partial_i\Phi
-
\mathcal{C} \partial_i\delta\phi\Big]
=0.
\end{equation}

The final conservation equations are as follows:
\begin{align}
\frac{\mathrm{d} \delta_m}{\mathrm{d} \ln a}&=-\frac{\theta_m}{\mathcal{H}}+ 3 \frac{\mathrm{d} \Phi}{\mathrm{d} \ln a},
\label{continuityeqlnafullSFC}
\\[6pt]
\frac{\mathrm{d}\theta_m}{\mathrm{d}\ln a}
={}&
-\left(
1
-
\mathcal{C}
\frac{\mathrm{d}\bar\phi}{\mathrm{d}\ln a}
\right)\theta_m+
\frac{k^2}{\mathcal H}
\left(
\Phi-\mathcal{C}\delta\phi
\right),
\label{eulereqlnafullSFC}\\[6pt]
\frac{\mathrm{d}^{2}\delta\phi}
{\mathrm{d}(\ln a)^{2}}
={}&
-\left(
2+\frac{\mathrm{d}\ln\mathcal{H}}
{\mathrm{d}\ln a}
\right)
\frac{\mathrm{d}\delta\phi}{\mathrm{d}\ln a}-
\left[
\frac{k^{2}}{\mathcal{H}^{2}}
+
\frac{\epsilon_d a^{2}}{\mathcal{H}^{2}}
\left(
\frac{\mathrm{d}^{2}V}{\mathrm{d}\bar{\phi}^{2}}
+\beta^{2}\bar{\rho}_{\mathrm{int}}
\right)
\right]\delta\phi+
\frac{\epsilon_d a^{2}\beta\bar{\rho}_{\mathrm{int}}}
{\mathcal{H}^{2}}\delta_m
\nonumber\\
&-
\frac{2\epsilon_d a^{2}\Phi}{\mathcal{H}^{2}}
\left(
\frac{\mathrm{d}V}{\mathrm{d}\bar{\phi}}
-\beta\bar{\rho}_{\mathrm{int}}
\right)+
4
\frac{\mathrm{d}\Phi}{\mathrm{d}\ln a}
\frac{\mathrm{d}\bar{\phi}}{\mathrm{d}\ln a}.
\label{KgperturbeqlnafullSFC}
\end{align}
where Eqs. \eqref{continuityeqlnafullSFC} and \eqref{KgperturbeqlnafullSFC} are obtained by substituting Eqs. \eqref{conservationkg} and \eqref{conservationeffmatter2} into the total conservation equation, Eq. \eqref{Eq:totalconservationInteracting}. 
Finally, Eq. \eqref{eulereqlnafullSFC} is obtained from using $\theta_m\equiv\partial_i v_m^i,$ and the Fourier form of Eq. \eqref{velocityeqinteracting}. In the resulting system, all the conformal time derivatives are expressed entirely in terms of derivatives with respect to $\ln a$. Therefore,  Eqs. \eqref{zerozeroeqlnafullSFC} and \eqref{zeroieqlnafullSFC} with \eqref{continuityeqlnafullSFC},\eqref{eulereqlnafullSFC} and \eqref{KgperturbeqlnafullSFC} give the full set of perturbation equations for the interacting scalar field dark energy model. Some subtleties related to the energy momentum conservation of the matter sector and the scalar field sector are discussed in detail in Appendix \ref{emcon}.

\begin{figure*}
\centering
\subfigure[Redshift evolution of $\%$ error $\mathcal{E}_{D}^{SHA}$ for SHA approximation in $D$ for $k=0.01\,\mathrm{Mpc}^{-1}$]
{\includegraphics[width=0.47\textwidth,height=52mm]{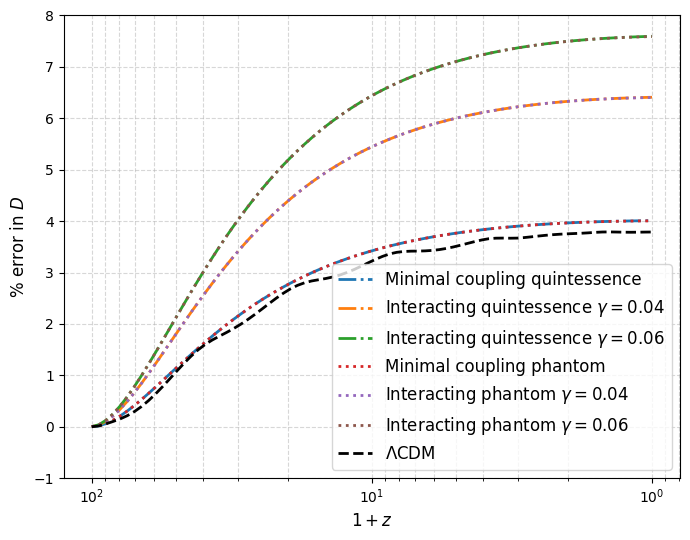}
\label{SHAerror_D_k_0.01}}
\hfill
\subfigure[Redshift evolution of $\%$ error for $\mathcal{E}_{f}^{SHA}$ SHA approximation in $f$ for $k=0.01\,\mathrm{Mpc}^{-1}$]
{\includegraphics[width=0.47\textwidth,height=52mm]{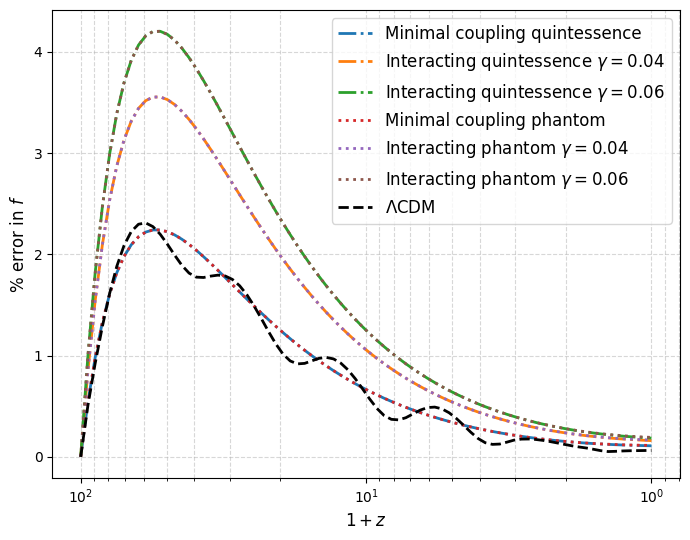}
\label{SHAerror_f_k_0.01}}

\subfigure[Redshift evolution of $\%$ error $\mathcal{E}_{D}^{SHA}$ for SHA approximation in $D$ for $k=0.005\,\mathrm{Mpc}^{-1}$]
{\includegraphics[width=0.47\textwidth,height=52mm]{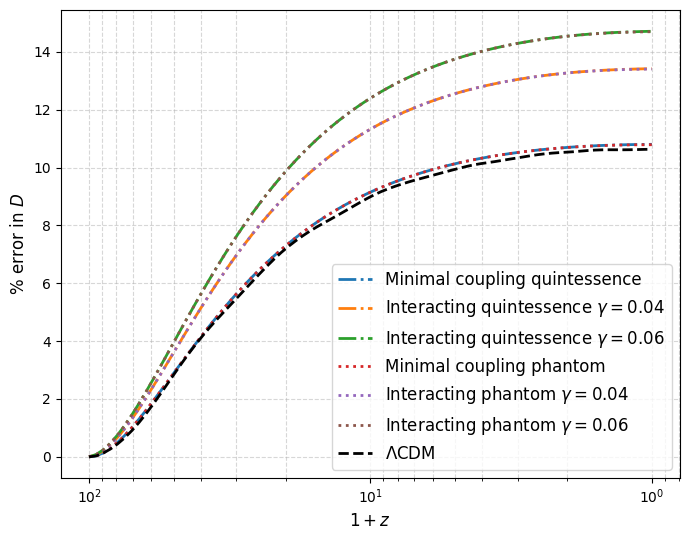}
\label{SHAerror_D_k_0.005}}
\hfill
\subfigure[Redshift evolution of $\%$ error $\mathcal{E}_{f}^{SHA}$ for SHA approximation in $f$ for $k=0.005\,\mathrm{Mpc}^{-1}$]
{\includegraphics[width=0.47\textwidth,height=52mm]{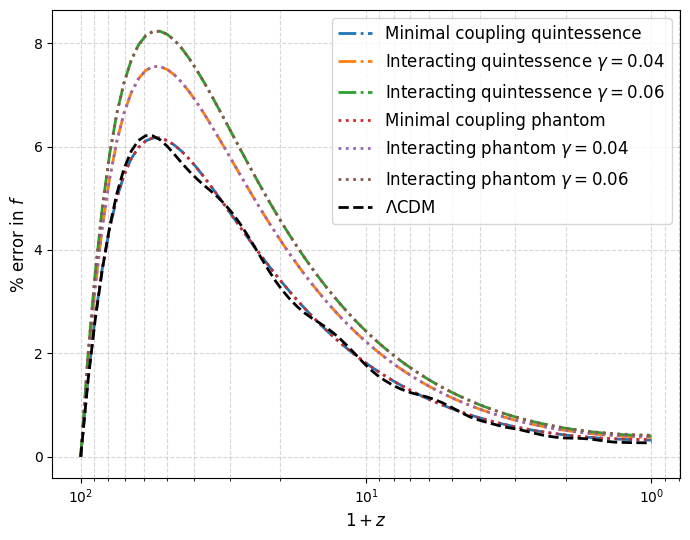}
\label{SHAerror_f_k_0.005}}

\subfigure[Redshift evolution of $\%$ error $\mathcal{E}_{D}^{SHA}$ for SHA approximation in $D$ for $k=0.003\,\mathrm{Mpc}^{-1}$]
{\includegraphics[width=0.47\textwidth,height=52mm]{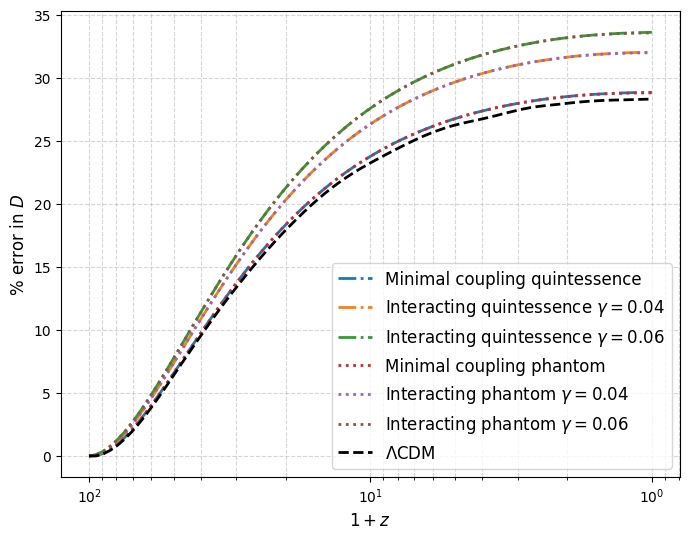}
\label{SHAerror_D_k_0.003}}
\hfill
\subfigure[Redshift evolution of $\%$ error $\mathcal{E}_{f}^{SHA}$ for SHA approximation in $f$ for $k=0.003\,\mathrm{Mpc}^{-1}$]
{\includegraphics[width=0.47\textwidth,height=52mm]{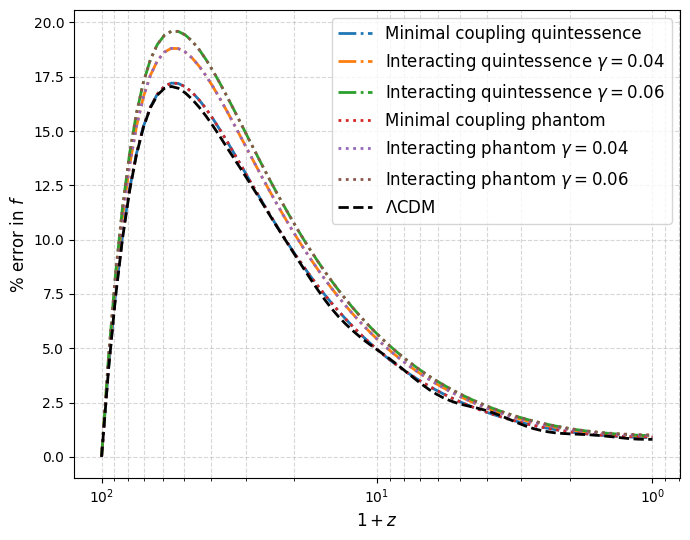}
\label{SHAerror_f_k_0.003}}

\caption{Redshift evolution of the percentage error $\mathcal{E}_{D}^{SHA}$ and $\mathcal{E}_{f}^{SHA}$ in the cold dark matter
linear growth factor,
$D(a,k)=\delta_m(a,k)/\delta_m(a_{\rm ini},k)$ (left panels), and the
corresponding linear growth rate,
$f(a,k)=\mathrm{d}\ln D/\mathrm{d}\ln a$ (right panels), introduced
by the sub-horizon approximation (SHA) using Eq. \eqref{errorfromfullsoln}. The errors are
evaluated for $\Lambda$CDM, the non-interacting scalar-field models,
and the interacting scalar-field models with two parameter sets:
(i) $(\gamma,\beta)=(0.04,0.001)$ and
(ii) $(\gamma,\beta)=(0.06,0.001)$. In each case, the percentage error
is computed relative to the corresponding full linear solution
obtained from the coupled $\delta$--$\theta$ system. Results are shown for three
representative comoving wavenumbers,
$k=0.01$, $0.005$, and $0.003\,\mathrm{Mpc}^{-1}$, for both
quintessence and phantom scalar-field dark energy. The approximation error is inversely correlated with the wavenumber $k$, and the overall amplitude of this error is highly model-dependent.}
\label{SHA_k_error}
\end{figure*}
\subsubsection{Sub-horizon Approximation}

Applying the sub-horizon approximation, $k\gg\mathcal{H}$, given by Eq.~(\ref{SHAcondition}) keeping only the zeroth-order terms $\mathcal{O}(\epsilon^0)$, and subsequently using the Einstein equations, Eqs.~(\ref{zerozeroeqlnafullSFC}--\ref{zeroieqlnafullSFC}), to simplify the perturbation equations, Eqs.~(\ref{continuityeqlnafullSFC}--\ref{KgperturbeqlnafullSFC}), we obtain, 
\begin{align}
-k^2 \Phi
={}&
\frac{3\mathcal{H}^2}{2}
\left[
\Omega_m\left(\delta_m
-\mathcal{C}\delta \phi\right)
+\Omega_\phi\delta_{\phi}
\right],\label{poissoneqlnaSHASFC}\\[6pt]
\frac{\mathrm{d}\delta_m}{\mathrm{d}\ln a}
={}&-\frac{\theta_m}{\mathcal{H}},
\label{continuityeqlnaSHASFC}\\[6pt]
\frac{\mathrm{d}\theta_m}{\mathrm{d}\ln a}
={}&
-\left(
1
-
\mathcal{C}
\frac{\mathrm{d}\bar\phi}{\mathrm{d}\ln a}
\right)\theta_m-
\frac{3\mathcal{H}}{2}
\left[
\Omega_m\left(\delta_m
-\mathcal{C}\delta \phi\right)
+\Omega_\phi\delta_{\phi}
\right]-\frac{\epsilon_d a^{2}\beta\mathcal{C}\bar{\rho}_{\mathrm{int}}\delta_m}{\mathcal{H}}.
\label{eulereqlnaSHASFC}\\[6pt]
\delta \phi =
{}&
\frac{\epsilon_d a^{2}\beta\bar{\rho}_{\mathrm{int}}\delta_m}{
k^{2}
+\epsilon_d a^{2}
\left(
\frac{\mathrm{d}^{2}V}{\mathrm{d}\bar{\phi}^{2}}
+\beta^{2}\bar{\rho}_{\mathrm{int}}
\right)}\label{KgeqlnaSHASFC}
\end{align}
\begin{figure*}
\subfigure[Redshift evolution of $\%$ error for QSA approximation in $D$ for $k=0.01Mpc^{-1}$]
{\includegraphics[width=0.47\textwidth,height=52mm]{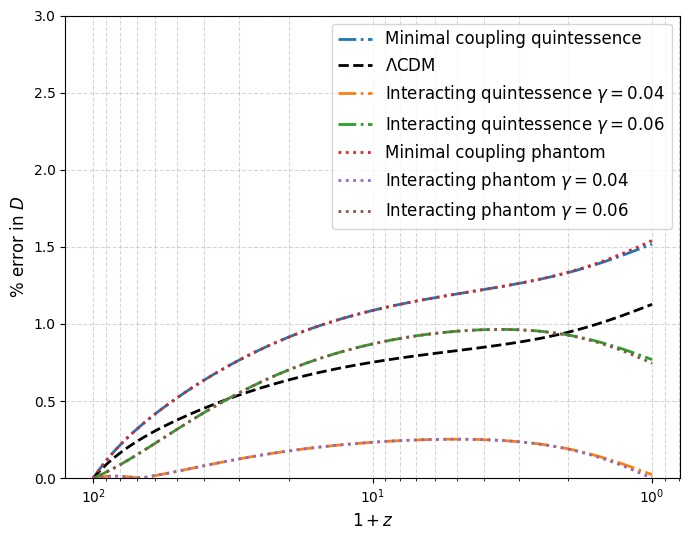}\label{QSAerror_D_k_0.01}}
\hspace{0.2cm}
\subfigure[Redshift evolution of $\%$ error for QSA approximation in $f$ for $k=0.01Mpc^{-1}$]
{\includegraphics[width=0.47\textwidth,height=52mm]{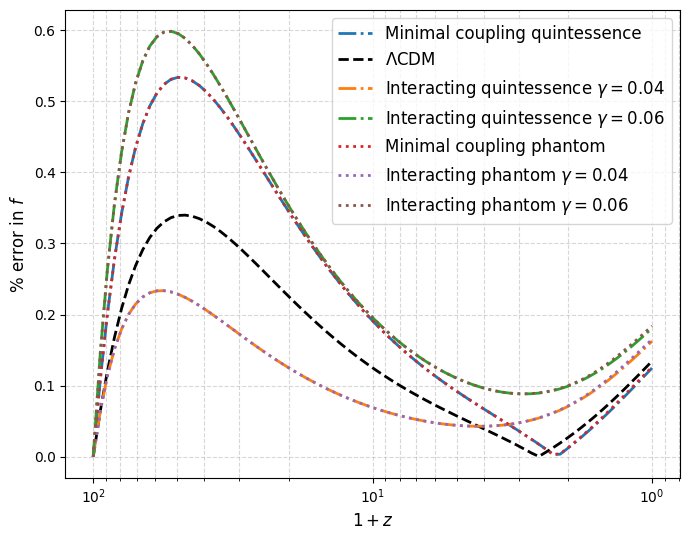}\label{QSAerror_f_k_0.01}}

\subfigure[Redshift evolution of $\%$ error for QSA approximation in $D$ for $k=0.005Mpc^{-1}$]
{\includegraphics[width=0.47\textwidth,height=52mm]{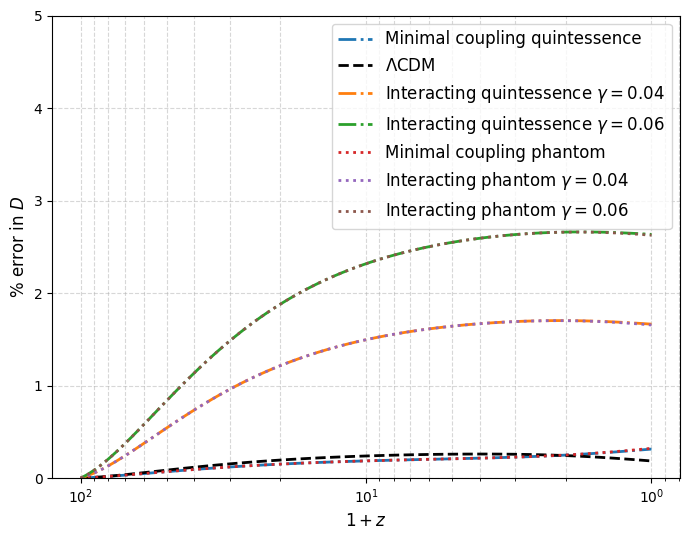}\label{QSAerror_D_k_0.005}}
\hspace{0.2cm}
\subfigure[Redshift evolution of $\%$ error for QSA approximation in $f$ for $k=0.005Mpc^{-1}$ ]
{\includegraphics[width=0.47\textwidth,height=52mm]{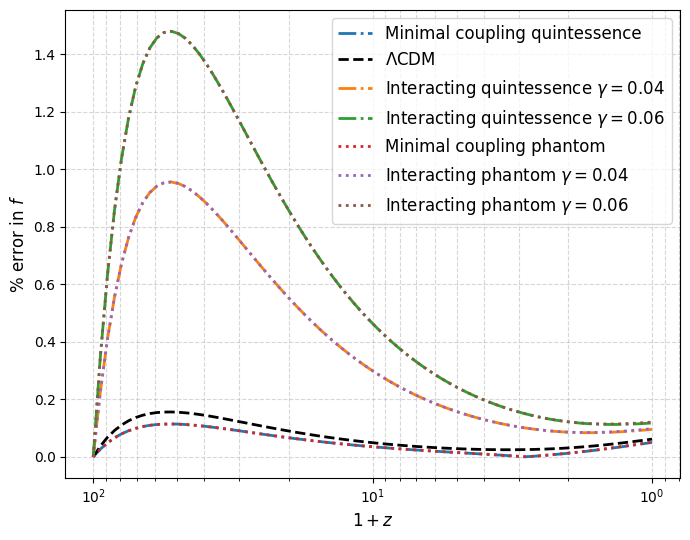}\label{QSAerror_f_k_0.005}}

\subfigure[Redshift evolution of $\%$ error for QSA approximation in $D$ for $k=0.003Mpc^{-1}$]
{\includegraphics[width=0.47\textwidth,height=52mm]{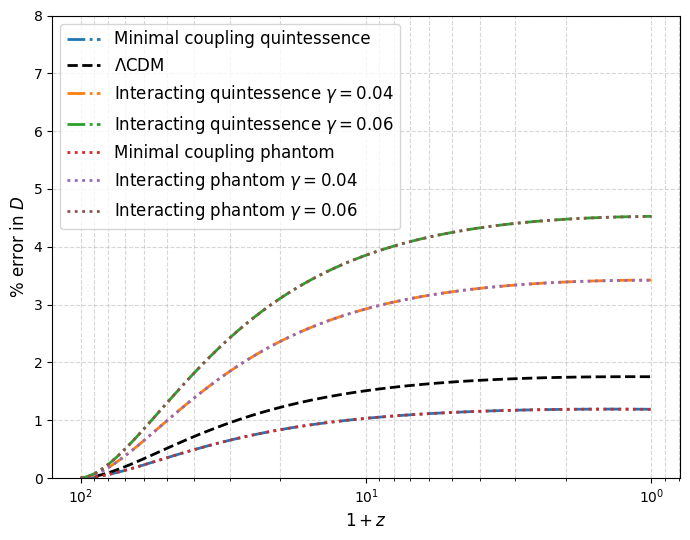}\label{QSAerror_D_k_0.003}}
\hspace{0.2cm}
\subfigure[Redshift evolution of $\%$ error for QSA approximation in $f$ for $k=0.003Mpc^{-1}$]
{\includegraphics[width=0.47\textwidth,height=52mm]{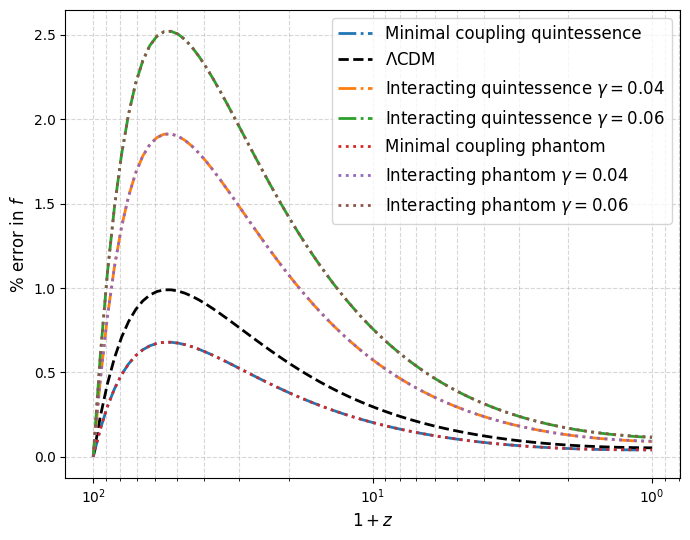}\label{QSAerror_f_k_0.003}}

\caption{Redshift evolution of the percentage error $\mathcal{E}_{D}^{QSA}$ and $\mathcal{E}_{f}^{QSA}$ in the cold dark matter
linear growth factor,
$D(a,k)=\delta_m(a,k)/\delta_m(a_{\rm ini},k)$ (left panels), and the
corresponding linear growth rate,
$f(a,k)=\mathrm{d}\ln D/\mathrm{d}\ln a$ (right panels), introduced
by the quasi-static approximation (QSA) using Eq. \eqref{errorfromfullsoln}. The errors are
evaluated for $\Lambda$CDM, the non-interacting scalar-field models,
and the interacting scalar-field models with two parameter sets:
(i) $(\gamma,\beta)=(0.04,0.001)$ and
(ii) $(\gamma,\beta)=(0.06,0.001)$. In each case, the percentage error
is computed relative to the corresponding full linear solution
obtained from the coupled $\delta$--$\theta$ system. Results are shown for three
representative comoving wavenumbers,
$k=0.01$, $0.005$, and $0.003\,\mathrm{Mpc}^{-1}$, for both
quintessence and phantom scalar-field dark energy. The approximation error is inversely dependent on the wavenumber $k$ and varies across different cosmological models; the QSA consistently yields a smaller error compared to the SHA.
}\label{QSA_k_error}
\end{figure*}
\begin{figure*}
\subfigure[$\%$ error $\mathcal{E}_{D}^{SHA}$ with $\gamma$ for SHA approximation in $D$ at $z=0$ for quintessence dark energy]
{\includegraphics[width=0.47\textwidth,height=52mm]{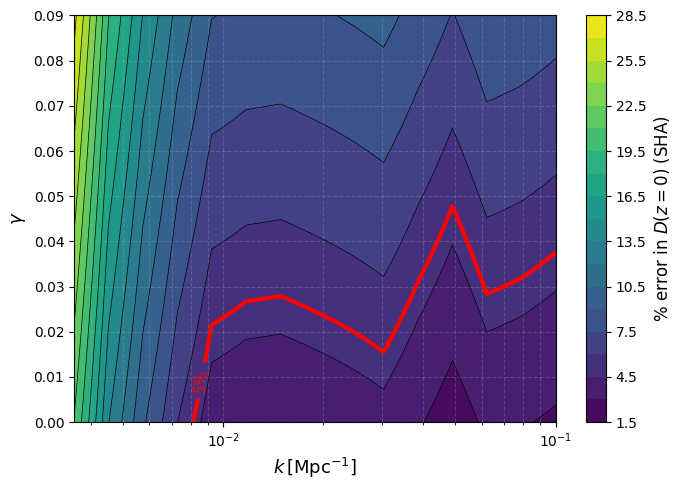}\label{SHA_error_D}}
\hspace{0.1cm}
\subfigure[$\%$ error $\mathcal{E}_{D}^{SHA}$ with $\gamma$ for SHA approximation in $D$ at $z=0$ for phantom dark energy]
{\includegraphics[width=0.47\textwidth,height=52mm]{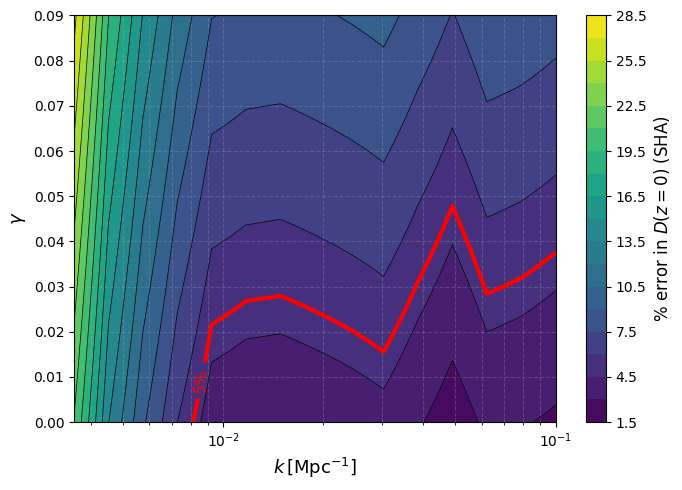}\label{SHA_error_D_phantom}}

\caption{Percentage error $\mathcal{E}_{D}^{SHA}$ of the sub-horizon approximation (SHA) relative to the full linear solution for the matter growth factor $D$ at $z=0$ using Eq. \eqref{errorfromfullsoln}, as the interaction parameter $\gamma$ is varied. The left and right panels correspond to quintessence and phantom scalar-field dark energy, respectively. The red contours indicate the $5\%$ error threshold relative to the full solution for $D$. The approximation error depends on the wavenumber $k$ and is also highly sensitive to the interaction parameter $\gamma$.}
\label{parameter_k_error}
\end{figure*}
\subsubsection{Quasi-static Approximation}

Applying the quasi-static approximation, Eq.~(\ref{QSAcondition}), and neglecting the higher order time derivatives of the gravitational potentials, keeping only the zeroth-order terms $\mathcal{O}(\epsilon_{\Phi}^0)$, and subsequently using Eq.~(\ref{zeroieqlnafullSFC}), to simplify the perturbation equations, Eq. \eqref{zerozeroeqlnafullSFC} and Eqs.~(\ref{continuityeqlnafullSFC}--\ref{KgperturbeqlnafullSFC}), we obtain,
\begin{align}
-k^{2}\Phi&=
\frac{3\mathcal{H}^2}{2}
\left[
\Omega_m\left(\delta_m
-\mathcal{C}\delta \phi\right)
+\Omega_\phi\delta_{\phi}
\right]+\frac{9}{2}\frac{\mathcal{H}^{3}}{k^{2}}
\Big(
\Omega_{m}{\theta}_{m}
+
\left(1+\omega_{\phi}\right)
\Omega_{\phi}{\theta}_{\phi}
\Big),
\label{zerozeroeqlnaQSASFC}\\[6pt]
\frac{\mathrm{d}\delta_m}{\mathrm{d}\ln a}
 &= -\frac{\theta_m}{\mathcal{H}}\label{continuityeqlnaQSASFC}\\[6pt]
\frac{\mathrm{d}{\theta}_{m}}{\mathrm{d}\ln a}
&=-\left(
1
-
\mathcal{C}
\frac{\mathrm{d}\bar\phi}{\mathrm{d}\ln a}
\right)\theta_m+
\frac{k^2}{\mathcal H}
\left(
\Phi-\mathcal{C}\delta\phi
\right),\nonumber\\
\label{eulereqlnaQSASFC}\\[6pt]
\frac{\mathrm{d}^{2}\delta\phi}
{\mathrm{d}(\ln a)^{2}}
={}&
-\left(
2+\frac{\mathrm{d}\ln\mathcal{H}}
{\mathrm{d}\ln a}
\right)
\frac{\mathrm{d}\delta\phi}{\mathrm{d}\ln a}-
\left[
\frac{k^{2}}{\mathcal{H}^{2}}
+
\frac{\epsilon_d a^{2}}{\mathcal{H}^{2}}
\left(
\frac{\mathrm{d}^{2}V}{\mathrm{d}\bar{\phi}^{2}}
+\beta^{2}\bar{\rho}_{\mathrm{int}}
\right)
\right]\delta\phi
+
\frac{\epsilon_d a^{2}\beta\bar{\rho}_{\mathrm{int}}}
{\mathcal{H}^{2}}\delta_m
\nonumber\\
&-
\frac{2\epsilon_d a^{2}\Phi}{\mathcal{H}^{2}}
\left(
\frac{\mathrm{d}V}{\mathrm{d}\bar{\phi}}
-\beta\bar{\rho}_{\mathrm{int}}
\right).
\label{KgperturbeqlnaQSASFC}
\end{align}
Eqs.~(\ref{zerozeroeqlnafullSFC}--\ref{zeroieqlnafullSFC}) and Eqs.~(\ref{continuityeqlnafullSFC}--\ref{KgperturbeqlnafullSFC}), together with the background evolution equations, form a closed system that fully describes the linear perturbation dynamics of scalar-field dark energy in the interacting model. The evolution of matter and scalar-field perturbations under the sub-horizon and quasi-static approximations is then obtained by solving Eqs.~(\ref{continuityeqlnaSHASFC})--(\ref{KgeqlnaSHASFC}) and Eqs.~(\ref{zerozeroeqlnaQSASFC})--(\ref{KgperturbeqlnaQSASFC}), respectively, together with the same background evolution equations.
\subsection{ Non-interacting scalar field dark energy}

For the non-interacting scalar-field dark energy model, the corresponding linear perturbation equations have been derived previously in the literature \cite{Unnikrishnan:2008qe, Ma:1994dv, Richarte:2016qqm}. So, we do not re-derive them separately here; instead, the full set of linearized perturbation equations for the non-interacting model can be obtained directly from the interacting case by setting the interaction parameter $\gamma=0$. In this limit, all interaction-dependent terms vanish, yielding the perturbation equations for the non-interacting scalar-field model. This limit also provides a consistency check of the interacting model, as the resulting equations recover the previously established perturbation equations for the non-interacting quintessence field.
\subsubsection{Full linearized perturbation equations}
The complete set of perturbation equations consists of two equations obtained from the linearized Einstein equations and three equations derived from the energy--momentum conservation equations, as in the previous interacting model; they can also be directly obtained by setting $\gamma=0$ in Eqs.~(\ref{zerozeroeqlnafullSFC}--\ref{zeroieqlnafullSFC}) and Eqs.~(\ref{continuityeqlnafullSFC}--\ref{KgperturbeqlnafullSFC}). 

The corresponding equations will be,
\begin{align}
-k^{2}\Phi
&= 3\mathcal{H}^{2}
\left(
\frac{\mathrm{d}\Phi}{\mathrm{d}\ln a} + \Phi
\right)
+
\frac{3\mathcal{H}^{2}}{2}
\Big(
\Omega_{m}\,\delta_{m}
+ \Omega_{\phi}\,\delta_{\phi}
\Big),
\label{zerozeroeqlnafullSF}
\\[6pt]
\frac{\mathrm{d}\Phi}{\mathrm{d}\ln a} + \Phi
&=
\frac{3\mathcal{H}}{2k^{2}}
\Big[
\Omega_{m}\,{\theta}_{m}
+
\left(1+\omega_{\phi}\right)
\Omega_{\phi}\,{\theta}_{\phi}
\Big],
\label{zeroieqlnafullSF}\\[6pt]
\frac{\mathrm{d}\delta_{m}}{\mathrm{d}\ln a}
+\frac{\theta_{m}}{\mathcal{H}}
&=+ 3\,\frac{\mathrm{d}\Phi}{\mathrm{d}\ln a},
\label{continuityeqlnafullSF}
\\[6pt]
\frac{\mathrm{d}{\theta}_{m}}{\mathrm{d}\ln a}
+ {\theta}_{m}
&=
\frac{k^{2}}{\mathcal{H}}\,\Phi,
\label{eulereqlnafullSF}\\[6pt]
\frac{\mathrm{d^{2}}\delta \phi}{\mathrm{d}\left(\ln a\right)^{2}}&=-\left(2 + \frac{\mathrm{d}\ln \mathcal{H}}{\mathrm{d}\ln a}\right)\frac{\mathrm{d}\delta \phi}{\mathrm{d}\ln a}-\left(\frac{k^{2}}{\mathcal{H}^{2}}+\frac{\epsilon_{d}a^{2}}{\mathcal{H}^{2}}\frac{\mathrm{d^{2}}V}{\mathrm{d}\bar{\phi}^{2}}\right)\delta\phi
-\frac{2\epsilon_{d}\Phi a^{2}}{\mathcal{H}^{2}}\frac{\mathrm{d}V}{\mathrm{d}\bar{\phi}} +4\frac{\mathrm{d}\Phi}{\mathrm{d}\ln a}\frac{\mathrm{d}\bar{\phi}}{\mathrm{d}\ln a},\nonumber\\
\label{KgperturbeqlnafullSF}
\end{align}
\subsubsection{Sub-horizon Approximation}

Following the same procedure as in the interacting model, we apply the sub-horizon approximation, $k\gg\mathcal{H}$, given by Eq.~\eqref{SHAcondition}, to the linearized perturbation equations Eqs.~\eqref{zerozeroeqlnafullSF}--\eqref{KgperturbeqlnafullSF}. Subsequently, using the Einstein equations, Eqs.~\eqref{zerozeroeqlnafullSF}--\eqref{zeroieqlnafullSF}, to eliminate the metric perturbations, we obtain the corresponding perturbation equations under the sub-horizon approximation. Equivalently, these equations are same as the SHA equations for the interacting scalar-field dark energy model, Eqs.~\eqref{poissoneqlnaSHASFC}--\eqref{eulereqlnaSHASFC}, since the interaction parameter $\gamma$ does not contribute to the perturbation equations at the SHA level.
\begin{align}
k^{2}\Phi
    &= - \frac{3\mathcal{H}^{2}}{2}\,\,
      \Omega_{m}\,\delta_{m}.\label{poissoneqlnaSHASF}\\[6pt]
\frac{\mathrm{d}\delta_{m}}{\mathrm{d}\ln a}
+\frac{\theta_{m}}{\mathcal{H}}
&=0,
\label{continuityeqlnaSHASF}\\[6pt]
\frac{\mathrm{d}{\theta}_{m}}{\mathrm{d}\ln a}
+
{\theta}_{m}
&=
-\frac{3\mathcal{H}}{2}
\Omega_{m}\,\delta_{m},
\label{eulereqlnaSHASF}
\end{align}
\subsubsection{Quasi-static Approximation}

Again, applying the quasi-static approximation, Eq.~\eqref{QSAcondition}, as discussed in the interacting dark sector model, and neglecting the time derivatives of the gravitational potentials, subsequently simplifying the perturbation equations using Eq.~\eqref{zeroieqlnafullSF} we get the QSA approximate equations. Equivalently, the same set of equations can be obtained from Eqs.~\eqref{zerozeroeqlnaQSASFC}--\eqref{KgperturbeqlnaQSASFC} by setting the interaction parameter $\gamma=0$. The resulting system consists of the following four equations:
\begin{align}
-k^{2}\Phi
&=
\frac{3\mathcal{H}^{2}}{2}
\left(
\Omega_{m}\delta_{m}
+
\Omega_{\phi}\delta_{\phi}
\right)+\frac{9}{2}\frac{\mathcal{H}^{3}}{k^{2}}
\left[
\Omega_{m}{\theta}_{m}
+
\left(1+\omega_{\phi}\right)
\Omega_{\phi}{\theta}_{\phi}
\right],
\label{poissoneqlnaQSASF}\\[6pt]
\frac{\mathrm{d}\delta_{m}}{\mathrm{d}\ln a}+\frac{\theta_{m}}{\mathcal{H}}
&=0
,
\label{continuityeqlnaQSASF}\\[6pt]
\frac{\mathrm{d}{\theta}_{m}}{\mathrm{d}\ln a}
+
{\theta}_{m}
&=
-\frac{3\mathcal{H}}{2}
\left(
\Omega_{m}\delta_{m}
+
\Omega_{\phi}\delta_{\phi}
\right)
-\frac{9}{2}\frac{\mathcal{H}^{2}}{k^{2}}
\left[
\Omega_{m}{\theta}_{m}
+
\left(1+\omega_{\phi}\right)
\Omega_{\phi}{\theta}_{\phi}
\right],
\label{eulereqlnaQSASF}\\[6pt]
\frac{\mathrm{d^{2}}\delta \phi}{\mathrm{d}\left(\ln a\right)^{2}}&=-\left(2 + \frac{\mathrm{d}\ln \mathcal{H}}{\mathrm{d}\ln a}\right)\frac{\mathrm{d}\delta \phi}{\mathrm{d}\ln a}-\left(\frac{k^{2}}{\mathcal{H}^{2}}+\frac{\epsilon_{d}a^{2}}{\mathcal{H}^{2}}\frac{\mathrm{d^{2}}V}{\mathrm{d}\bar{\phi}^{2}}\right)\delta\phi
-\frac{2\epsilon_{d}\Phi a^{2}}{\mathcal{H}^{2}}\frac{\mathrm{d}V}{\mathrm{d}\bar{\phi}},
\label{KgperturbeqlnaQSASF}
\end{align}
Now, Eqs.~(\ref{zerozeroeqlnafullSF})--(\ref{KgperturbeqlnafullSF}), together with the background evolution equations, form a closed system that fully describes the linear perturbation dynamics of scalar-field dark energy in the non-interacting model. The corresponding perturbation evolution under the sub-horizon and quasi-static approximations is then obtained by solving Eqs.~(\ref{continuityeqlnaSHASF})--(\ref{eulereqlnaSHASF}) and Eqs.~(\ref{poissoneqlnaQSASF})--(\ref{KgperturbeqlnaQSASF}), respectively, supplemented by the same background evolution equations.
\subsection{$\Lambda$CDM}
For the $\Lambda$CDM model, the linearized perturbation equations are again well established in the literature and standard cosmology textbooks \cite{Baumann:2022mni}, and therefore we do not re-derive them here. The $\Lambda$CDM equations can instead be recovered from the scalar-field models by noting that the cosmological constant does not possess dynamical perturbations, i.e., $\delta_{\Lambda}=0$. In particular, starting from the interacting scalar-field model, taking the non-interacting limit $\gamma=0$ and setting the scalar-field perturbation $\delta_{\phi}=0$ removes both the interaction and the dynamical dark-energy perturbation, yielding the standard $\Lambda$CDM perturbation equations. Similarly, setting $\delta_{\phi}=0$ in the non-interacting scalar-field model recovers the same $\Lambda$CDM equations. Thus, the $\Lambda$CDM model is consistently recovered as the limit in which the dark-energy sector is reduced to the cosmological constant level with no perturbations.
\subsubsection{Full linearized perturbation equations}
In the $\Lambda$CDM model, the dark energy component is described by a cosmological constant and therefore does not possess an independent dynamical perturbation. Consequently, the full linearized perturbation system consists of the $00$ and $0i$ components of the Einstein equations together with the continuity and Euler equations for the cold dark matter.
\begin{align}
    -k^{2}\Phi
    - 3\mathcal{H}^{2}
    \left(
        \frac{\mathrm{d}\Phi}{\mathrm{d}\ln a} + \Phi
    \right)
    &= \frac{3\mathcal{H}^{2}}{2}\,\Omega_{m}\,\delta_{m},
    \label{zerozeroeqlnafullL} \\[6pt]
    \frac{\mathrm{d}\Phi}{\mathrm{d}\ln a} + \Phi
    &= \frac{3\mathcal{H}}{2k^{2}}\,
    \Omega_{m}\,\theta_{m},
    \label{zeroieqlnafullL}\\[6pt]
    \frac{\mathrm{d}\delta_{m}}{\mathrm{d}\ln a}
    &= -\frac{{\theta}_{m}}{\mathcal{H}}
    + 3\,\frac{\mathrm{d}\Phi}{\mathrm{d}\ln a},
    \label{continuityeqlnafullL} \\[6pt]
    \frac{\mathrm{d}{\theta}_{m}}{\mathrm{d}\ln a}
    + \theta_{m}
    &= \frac{k^{2}}{\mathcal{H}}\,\Phi,
    \label{eulereqlnafullL}
\end{align}
\begin{figure*}
\subfigure[Redshift evolution of $\Delta D^{A}$, the percentage deviation of $D$ of the quintessence scalar-field dark energy model from the $\Lambda$CDM solution at $k=0.01\,\mathrm{Mpc}^{-1}$.]
{\includegraphics[width=0.47\textwidth,height=49mm]{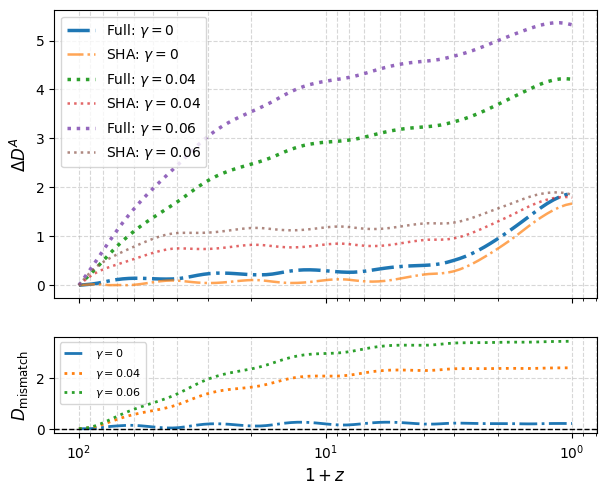}\label{diff_lcdm_D_k_0.01}}
\hspace{0.2cm}
\subfigure[Redshift evolution of $\Delta D^{A}$, the percentage deviation of $D$ of the phantom scalar-field dark energy model from the $\Lambda$CDM solution at $k=0.01\,\mathrm{Mpc}^{-1}$.]
{\includegraphics[width=0.47\textwidth,height=49mm]{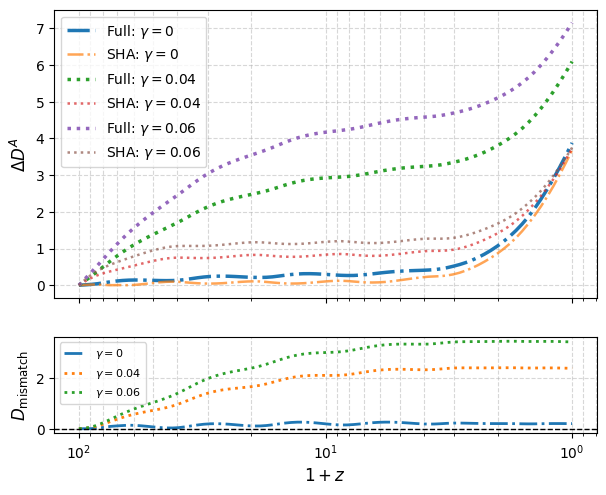}\label{diff_lcdm_D_phantom_k_0.01}}

\subfigure[Redshift evolution of $\Delta D^{A}$, the percentage deviation of $D$ of the quintessence scalar-field dark energy model from the $\Lambda$CDM solution at $k=0.005\,\mathrm{Mpc}^{-1}$.]
{\includegraphics[width=0.47\textwidth,height=49mm]{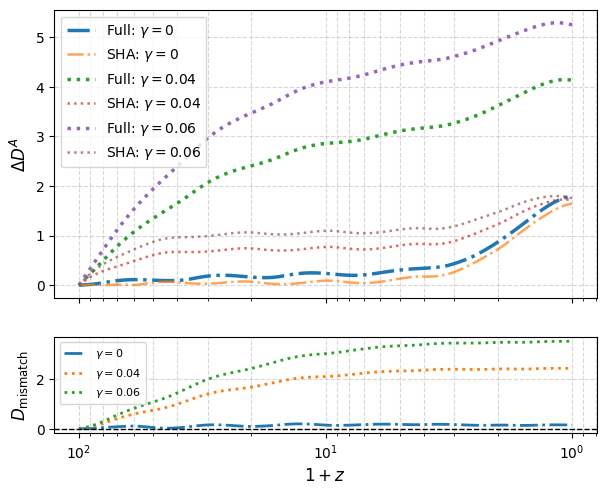}\label{diff_lcdm_D_k_0.005}}
\hspace{0.2cm}
\subfigure[Redshift evolution of $\Delta D^{A}$, the percentage deviation of $D$ of the phantom scalar-field dark energy model from the $\Lambda$CDM solution at $k=0.005\,\mathrm{Mpc}^{-1}$. ]
{\includegraphics[width=0.47\textwidth,height=49mm]{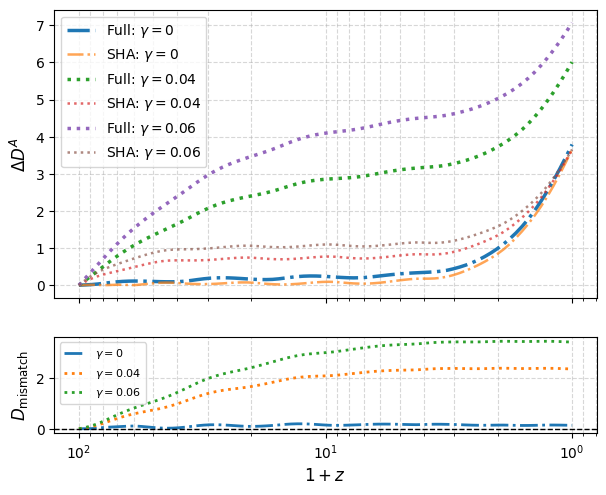}\label{diff_lcdm_D_phantom_k_0.005}}

\subfigure[Redshift evolution of $\Delta D^{A}$, the percentage deviation of $D$ of the quintessence scalar-field dark energy model from the $\Lambda$CDM solution at $k=0.003\,\mathrm{Mpc}^{-1}$.]
{\includegraphics[width=0.47\textwidth,height=49mm]{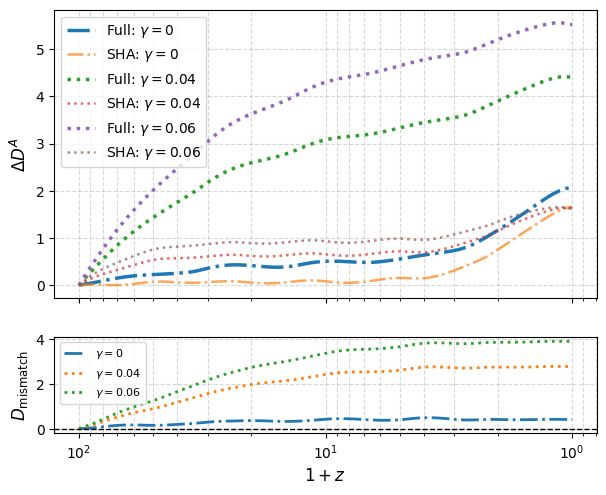}\label{diff_lcdm_D_k_0.003}}
\hspace{0.2cm}
\subfigure[Redshift evolution of $\Delta D^{A}$, the percentage deviation of $D$ of the phantom scalar-field dark energy model from the $\Lambda$CDM solution at $k=0.003\,\mathrm{Mpc}^{-1}$.]
{\includegraphics[width=0.47\textwidth,height=49mm]{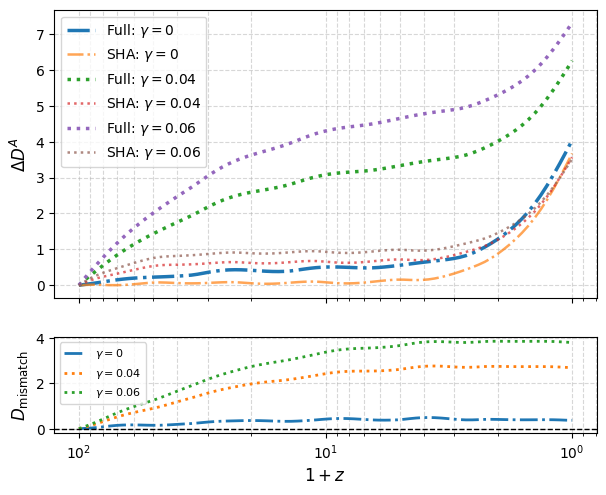}\label{diff_lcdm_D_phantom_k_0.003}}

\caption{Redshift evolution of $\Delta D^{A}$, the percentage deviation of the cold
dark matter linear growth factor,
$D(a,k)=\delta_m(a,k)/\delta_m(a_{\rm ini},k)$, from the corresponding
$\Lambda$CDM solution for quintessence (left panels) and phantom (right
panels) scalar-field dark energy. The top panels show the deviations
obtained using the full--full and SHA--SHA comparisons with $\Lambda$CDM model using Eq. \eqref{percentage_deviation},
the bottom panels show the difference between these deviations $D_{\rm{mismatch}}$, using Eq. \eqref{difference_deviation}. Results are
shown for $k=0.01$, $0.005$, and $0.003\,\mathrm{Mpc}^{-1}$ for the
non-interacting model ($\gamma=0$) and the interacting models with
$(\gamma,\beta)=(0.04,0.001)$ and $(0.06,0.001)$. $D_{\rm{mismatch}}$ exhibits a very weak inverse dependence on the wavenumber $k$.
}\label{diff_lcdm}
\end{figure*}
\begin{figure*}
\subfigure[Difference between the full--full and SHA--SHA percentage
deviations of the cold dark matter linear growth factor from the full
$\Lambda$CDM solution for the interacting quintessence
scalar-field dark energy model at $z=0$.]
{\includegraphics[width=0.47\textwidth,height=55mm]{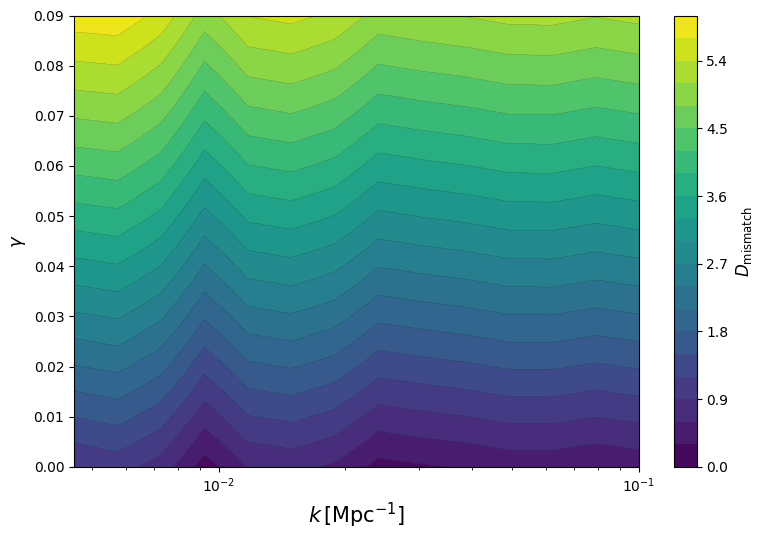}\label{deviation_quintessence}}
\hspace{0.1cm}
\subfigure[Difference between the full--full and SHA--SHA percentage
deviations of the cold dark matter linear growth factor from the full
$\Lambda$CDM solution for the interacting phantom
scalar-field dark energy model at $z=0$.]
{\includegraphics[width=0.47\textwidth,height=55mm]{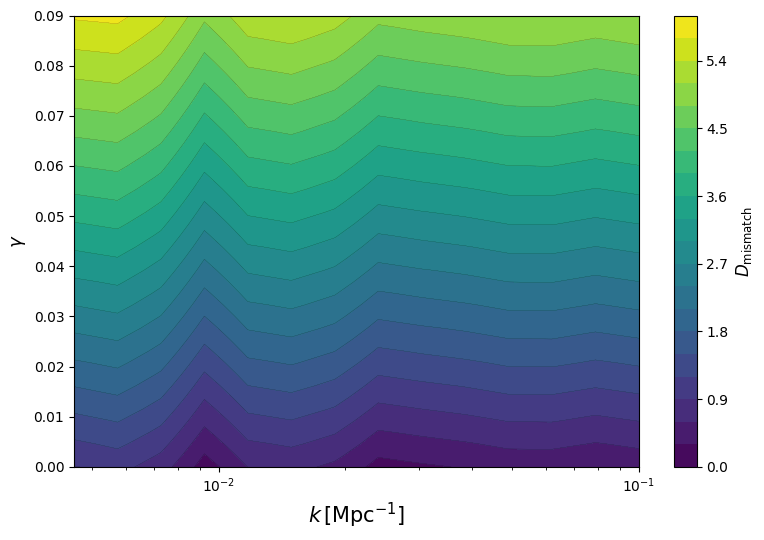}\label{deviation_phantom}}
\caption{Difference between the percentage deviations of the full--full
and SHA--SHA comparisons, defined by Eq.~\eqref{percentage_deviation},
for the cold dark matter linear growth factor relative to $\Lambda$CDM
at $z=0$. Here, full--full denotes the comparison of the full
scalar-field solution with the full $\Lambda$CDM solution, while
SHA--SHA denotes the comparison of the SHA scalar-field solution with
the SHA $\Lambda$CDM solution. The results are shown as a function of
the comoving wavenumber $k$ and interaction parameter
$\gamma\in[0,0.09]$, with all other model parameters held fixed. The
left and right panels correspond to quintessence and phantom
scalar-field dark energy, respectively. The non-interacting case is
included at $\gamma=0$. While $D_{\rm{mismatch}}$ exhibits a very weak inverse dependence on the wavenumber $k$, it is highly sensitive to the interaction parameter $\gamma$.}\label{deviation}
\end{figure*}
\subsubsection{Sub-horizon approximation}
In analogy with the previous two cases, we apply the sub-horizon approximation, Eq.~(\ref{SHAcondition}), and use the linearized Einstein equations, Eqs.~(\ref{zerozeroeqlnafullL}--\ref{zeroieqlnafullL}), to simplify the matter perturbation equations.
\begin{align}
k^{2}\Phi
    &= - \frac{3\mathcal{H}^{2}}{2}\,\,
      \Omega_{m}\,\delta_{m}.\label{poissoneqlnaSHAL}\\[6pt]
\frac{\mathrm{d}\delta_{m}}{\mathrm{d}\ln a}
    &= -\frac{{\theta}_{m}}{\mathcal{H}},
    \label{continuityeqlnaSHAL} \\[6pt]
    \frac{\mathrm{d}{\theta}_{m}}{\mathrm{d}\ln a}
    + {\theta}_{m}
    &= -\frac{3\mathcal{H}}{2}\,
      \Omega_{m}\,\delta_{m},
    \label{eulereqlnaSHAL}
\end{align}
\subsubsection{Quasi-static Approximation}
In the same manner, we apply the quasi-static approximation, Eq.~(\ref{QSAcondition}), and use the $00$ and $0i$ components of the linearized Einstein equations Eqs.~(\ref{zerozeroeqlnafullL}--\ref{zeroieqlnafullL}), to simplify the matter perturbation equations.
\begin{align}
k^{2}\Phi
&= - \frac{3\mathcal{H}^{2}}{2}\,
\Omega_{m}\,\delta_{m}
- \frac{9\mathcal{H}^{3}}{2k^{2}}\,
\Omega_{m}\,{\theta}_{m}, \label{poissoneqlnaQSAL}\\[6pt]
\frac{\mathrm{d}\delta_{m}}{\mathrm{d}\ln a}
&= -\frac{{\theta}_{m}}{\mathcal{H}}.
\label{continuityeqlnaQSAL}\\[6pt]
\frac{\mathrm{d}{\theta}_{m}}{\mathrm{d}\ln a}
+ {\theta}_{m}
&= - \frac{3\mathcal{H}}{2}\,\Omega_{m}\,\delta_{m}
- \,\frac{9\mathcal{H}^{2}}{2k^{2}}\,
\Omega_{m}\,{\theta}_{m},
\label{eulereqlnaQSAL}
\end{align}
The full linear evolution is obtained from Eqs.~\eqref{zerozeroeqlnafullL}--\eqref{eulereqlnafullL}, while the SHA and QSA evolutions of $\delta_m$ and $\theta_m$ follow from Eqs.~\eqref{continuityeqlnaSHAL}--\eqref{eulereqlnaSHAL} and Eqs.~\eqref{continuityeqlnaQSAL}--\eqref{eulereqlnaQSAL}, respectively, with the corresponding background evolution.
\section{Methodology}\label{Methodology}
\label{method}

To assess the validity of the sub-horizon and quasi-static approximations, we study the redshift evolution of cosmological perturbations by solving the full set of linear perturbation equations and comparing them with their SHA and QSA counterparts. The perturbation initial conditions are obtained in a self-consistent manner using the Boltzmann code \texttt{CLASS} \cite{DiegoBlas_2011}. Radiation is neglected in the analytical and numerical evolution considered here. We therefore choose the initial scale factor
$a_{\mathrm{ini}}=10^{-2}$, corresponding to an epoch at which the radiation contribution to the total energy density is already negligible.
\subsection*{Scalar-field dark energy}

For the scalar-field dark energy models, we first determine the background evolution, since the dark energy dynamics are governed by the scalar-field equation of motion. We construct both the non-interacting and interacting models such that their background expansion history matches that of the $\Lambda$CDM model within $1\%$, as shown in Fig.~(\ref{H_match_LCDM}). The interacting and non-interacting models yield approximately the same expansion history for the set of parameters chosen. This choice of parameters helps us to keep the background evolution approximately the same so that we can properly compare the perturbations on the various background evolutions. Therefore, their background evolution is presented together under the label scalar-field dark energy. For the non-interacting model, this gives,
\[
V_{0}=1.24\times10^{-7},
\qquad
\lambda=0.16,
\]
with the initial field values,
\[
\bar{\phi}=20,
\qquad
\bar{\phi}'=0.0.
\]
For the interacting model, the same scalar-field and potential parameters are retained, while the interaction is introduced through the parameters $\beta$ and $\gamma$. This specific interaction is not available in \texttt{CLASS}, so we modify the background and perturbation equations in the code accordingly and consider two representative parameter sets,
\[
(\beta,\gamma)=(0.001,0.04)
\]
and
\[
(\beta,\gamma)=(0.001,0.06).
\]
Using these inputs, \texttt{CLASS} evolves the scalar-field background and perturbations from the initial epoch to the present time. Using the output of \texttt{CLASS}, at the reference scale factor $a_{\mathrm{ini}}=10^{-2}$, we extract the background field $\bar{\phi}$ and its conformal-time derivative $\bar{\phi}'$, together with the perturbed  quantities $\delta_m$, $\theta_m$, $\delta_\phi$, and $\theta_\phi$. Since our background equations are formulated in terms of $\mathrm{d}\bar{\phi}/\mathrm{d}\ln a$, while \texttt{CLASS} provides $\bar{\phi}'$, we use
\begin{equation}
\bar{\phi}'=
\mathcal{H}\frac{\mathrm{d}\bar{\phi}}{\mathrm{d}\ln a}
\end{equation}
to obtain the corresponding logarithmic derivative at $a_{\mathrm{ini}}$. The resulting $\bar{\phi}$ and $\mathrm{d}\bar{\phi}/\mathrm{d}\ln a$ are then used as the initial conditions for the numerical integration of the background equations derived in this work.

The background equations are integrated in $\ln a$ using the adaptive fourth-order Runge--Kutta method (RK45), implemented through the \texttt{solve\_ivp} routine of the \texttt{SciPy} library \cite{2020SciPy-NMeth}. The integration is performed from $a_{\mathrm{ini}}$ to $a=1$, with the solution evaluated on a uniform grid in $\ln a$. The resulting scalar-field background quantities are interpolated using \texttt{InterpolatedUnivariateSpline} and subsequently supplied to the perturbation equations at the intermediate values of $a$ required by the adaptive integration.

For the perturbation evolution, the scalar-field perturbations provided by \texttt{CLASS} are converted into the field perturbation $\delta\phi$ and its conformal-time derivative $\delta\phi'$ using Eqs.~\eqref{theta_phi} and \eqref{delta_rho_phi}, respectively, as required by the linearized Klein--Gordon equation. Now, all the initial conditions are then used to integrate the full linearized perturbation equations and their SHA and QSA counterparts from $a_{\mathrm{ini}}$ to $a=1$ using the same adaptive RK45 scheme. The perturbation equations are solved separately for each selected comoving wavenumber $k$, with the integration performed in $\ln a$. For each $k$-mode, the solutions are evaluated on the chosen $\ln a$ output grid, allowing the scale dependence of the full, SHA, and QSA solutions to be compared.
\subsection*{$\Lambda$CDM}

For the $\Lambda$CDM model, the dark energy component is a cosmological constant and therefore has neither a dynamical scalar-field background nor an independent perturbation. Consequently, only the CDM perturbations, $\delta_m$ and $\theta_m$, are required for the numerical evolution. Their initial values at $a_{\mathrm{ini}}$ are extracted from the scalar perturbation output of \texttt{CLASS} using the $\Lambda$CDM model.

For each selected comoving wavenumber $k$, the full $\Lambda$CDM perturbation equations and their SHA and QSA counterparts are integrated in $\ln a$ from $a_{\mathrm{ini}}$ to $a=1$ using the adaptive RK45 scheme. The same initial conditions are used for all three systems, allowing a direct comparison between the full and approximate treatments.

As a consistency check of the numerical implementation, the solution of the full perturbation equations is compared with the corresponding \texttt{CLASS} evolution for both the $\Lambda$CDM and non-interacting scalar-field models. Excellent agreement to within $0.1\%$ is found in both cases, confirming the consistency of the numerical implementation and the perturbation initial conditions obtained from \texttt{CLASS}. Throughout the work, the numerical calculations and visualization are performed using \texttt{NumPy} \cite{harris2020array} and \texttt{Matplotlib} \cite{Hunter:2007}.
\begin{figure*}
\subfigure[Percentage deviation in the power spectrum relative to the full $\Lambda$CDM
solution at $z=0$ for quintessence scalar-field dark energy.]
{\includegraphics[width=0.47\textwidth,height=65mm]{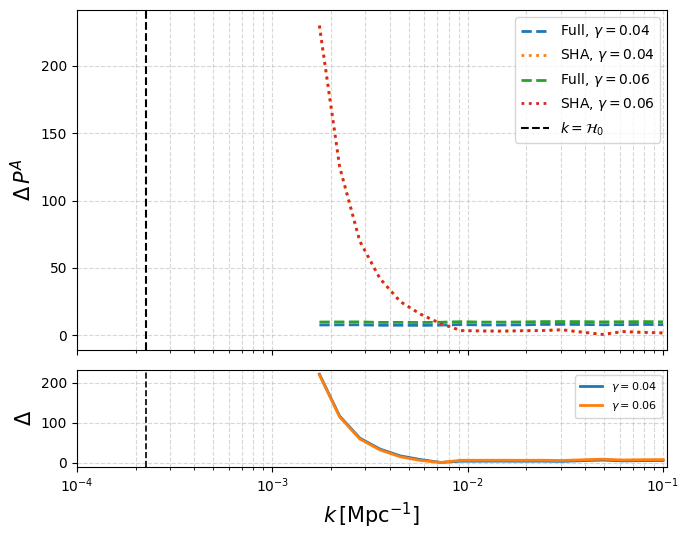}\label{percentage_error_P_full}}
\hspace{0.1cm}
\subfigure[Percentage deviation in the power spectrum relative to the full $\Lambda$CDM
solution at $z=0$ for phantom scalar-field dark energy.]
{\includegraphics[width=0.47\textwidth,height=65mm]{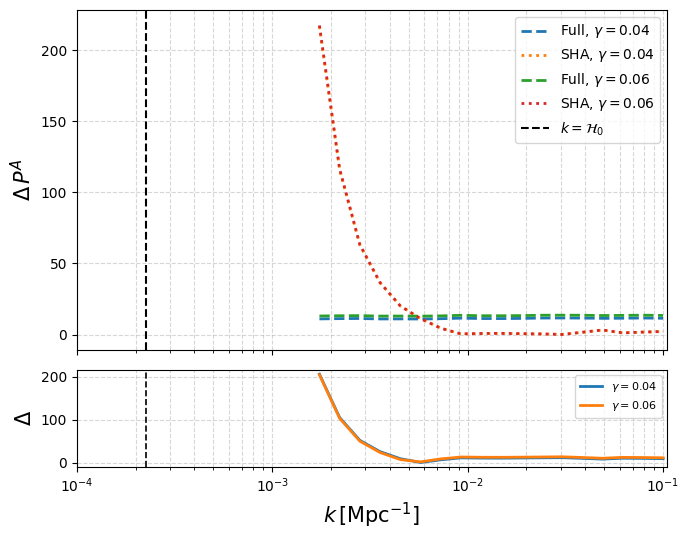}\label{percentage_error_P_full_phantom}}
\caption{Percentage deviations $\Delta P^{A}$ at $z=0$ of the full and SHA power spectra from the full $\Lambda$CDM power spectrum for the interacting scalar-field dark energy models with $(\gamma,\beta)=(0.04,0.001)$ and $(0.06,0.001)$, shown in the top panels using Eq. \eqref{powerspectrawrtlcdm}. The bottom panels show the difference between the SHA and full deviations from $\Lambda$CDM $\Delta$ using Eq. \eqref{powerspectradeviation}. The black dashed line marks the horizon scale, $k=\mathcal{H}$, at $z=0$. The left and right columns correspond to quintessence and phantom scalar-field dark energy, respectively.}\label{percentage_error_P_full}
\end{figure*}
\begin{figure*}
\subfigure[Difference between the deviations of the SHA and full power spectra
from the full $\Lambda$CDM power spectrum for quintessence scalar-field
dark energy.]
{\includegraphics[width=0.47\textwidth,height=55mm]{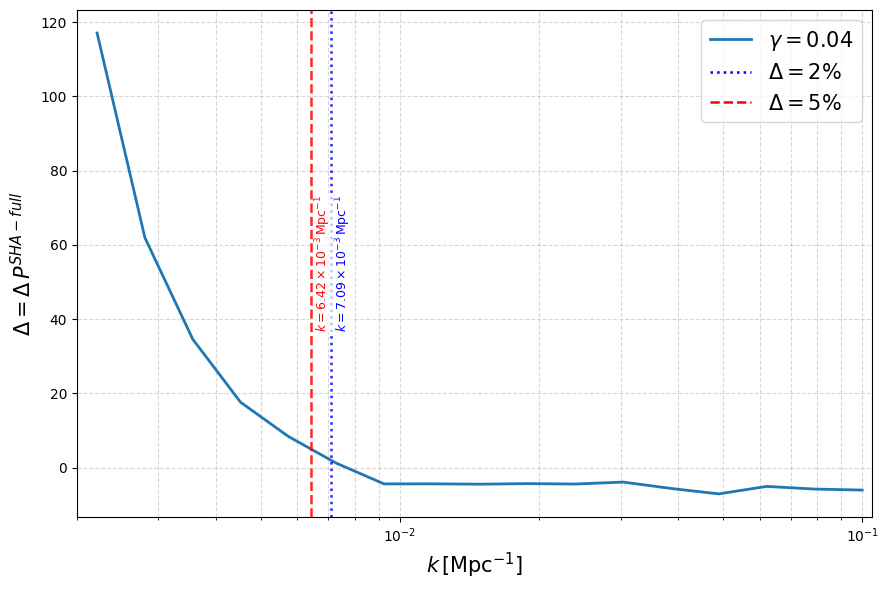}\label{diff_quintessence}}
\hspace{0.1cm}
\subfigure[Difference between the deviations of the SHA and full power spectra
from the full $\Lambda$CDM power spectrum for phantom scalar-field
dark energy.]
{\includegraphics[width=0.47\textwidth,height=55mm]{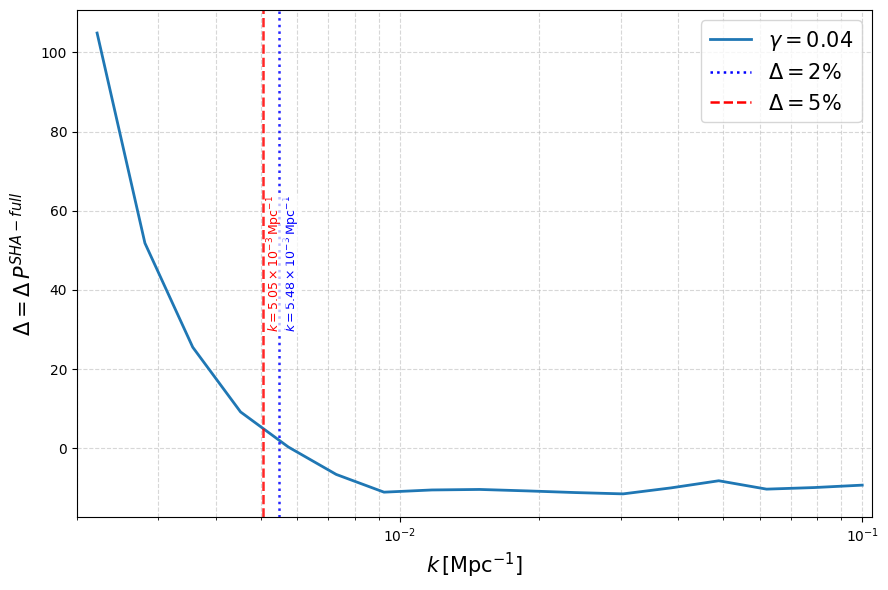}\label{diff_phantom}}
\caption{Difference $\Delta$ between the deviations of the SHA and full power spectra
from the full $\Lambda$CDM solution for the interacting scalar-field
dark energy model with $(\gamma,\beta)=(0.04,0.001)$ as a function of
the co-moving wavenumber $k$ at $z=0$ using Eq. \eqref{powerspectradeviation}. The blue and red dotted lines indicate the wavenumbers at which the SHA--Full difference reaches $2\%$ and $5\%$, respectively. The left
and right columns correspond to quintessence and phantom scalar-field
dark energy, respectively. The critical wavenumber $k$ associated with a specific percentage deviation is highly model-dependent.}\label{diff}
\end{figure*}
\subsection*{Analysis}

We quantify the accuracy of the sub-horizon approximation (SHA) and
quasi-static approximation (QSA) relative to the full linear evolution
using the matter density contrast. For each comoving wavenumber $k$,
the linear growth factor and growth rate are defined by,
\begin{equation}
D(a,k)=
\frac{\delta_m(a,k)}
{\delta_m(a_{\mathrm{ini}},k)},
\qquad
f(a,k)=
\frac{\mathrm{d}\ln D(a,k)}
{\mathrm{d}\ln a}.
\end{equation}
These quantities are obtained from the full, SHA, and QSA solutions.
Since the perturbation equations are solved independently in Fourier
space for each $k$, the linear matter power spectrum is obtained as,
\begin{equation}
P_m(k,a)
=
D^2(k,a)P_m(k,a_{\mathrm{ini}})
=
\left|
\frac{\delta_m(k,a)}
{\delta_m(k,a_{\mathrm{ini}})}
\right|^2
P_m(k,a_{\mathrm{ini}}),
\end{equation}
where $P_m(k,a_{\mathrm{ini}})$ is the common initial matter power
spectrum.

For $A=\mathrm{SHA},\mathrm{QSA}$, the accuracy of the approximate
solutions is quantified by their percentage deviations from the full
solution,
\begin{equation}\label{errorfromfullsoln}
\mathcal{E}_{X}^{A}(a,k)
=
100\left|
\frac{X^{A}(a,k)-X^{\mathrm{full}}(a,k)}
{X^{\mathrm{full}}(a,k)}
\right|,
\qquad
X=D,f.
\end{equation}
The redshift evolution of these errors is examined at
$k=0.01,\ 0.005,\ 0.003~\mathrm{Mpc}^{-1}$, while the dependence of the approximation accuracy on the interaction parameters is studied at $z=0$ over a $(k,\gamma)$ parameter space, sampled using 30 logarithmically spaced values in the range $k \in [10^{-4}, 10^{-1}]$ and 40 linearly spaced values in the range $\gamma \in [0, 0.09]$

To quantify the difference between the scalar-field models and
$\Lambda$CDM consistently within the same approximation scheme, we
compare the full scalar-field solutions with the full $\Lambda$CDM
solution, and the SHA scalar-field solutions with the SHA $\Lambda$CDM
solution. The percentage deviation is defined as,
\begin{equation}\label{percentage_deviation}
\Delta D^{A}
=
100\left|
\frac{
D^{A}-D_{\Lambda{\rm CDM}}^{A}
}{
D_{\Lambda{\rm CDM}}^{A}
}
\right|,
\qquad
A=\mathrm{full},\mathrm{SHA},
\end{equation}
where $D^{A}$ denotes the corresponding solution for the
non-interacting or interacting scalar-field model. Thus, $A=\mathrm{full}$
corresponds to a full--full comparison, while $A=\mathrm{SHA}$
corresponds to a SHA--SHA comparison. These deviations are evaluated
for both quintessence and phantom scalar-field dark energy at the
representative comoving wavenumbers considered above.

To quantify the change in the model deviation introduced by the SHA,
we further calculate the difference between the SHA and full
deviations,
\begin{equation}\label{difference_deviation}
D_{\rm{mismatch}}
=
|\Delta D^{\mathrm{SHA}}
-
\Delta D^{\mathrm{full}}|,
\end{equation}

We similarly investigate the impact of the approximations on the
matter power spectrum at $z=0$. The percentage deviation of the full
and SHA power spectra from the full $\Lambda$CDM spectrum is defined
as,
\begin{equation}\label{powerspectrawrtlcdm}
\Delta P^{A}(k)
=
100\left|
\frac{P^A(k)-P_{\Lambda{\rm CDM}}^{\mathrm{full}}(k)}
{P_{\Lambda{\rm CDM}}^{\mathrm{full}}(k)}
\right|,
\qquad
A=\mathrm{full},\mathrm{SHA}.
\end{equation}
This is evaluated for the non-interacting model and the two interacting
models, for both quintessence and phantom dark energy. To isolate the
additional effect of the SHA, we consider the difference between the
SHA and full deviations,
\begin{equation}\label{powerspectradeviation}
\Delta=
\Delta P^{\mathrm{SHA-full}}(k)
=
\Delta P^{\mathrm{SHA}}(k)
-
\Delta P^{\mathrm{full}}(k),
\end{equation}
\section{Results}
\label{result}

In the sub-horizon approximation (SHA) of $\Lambda$CDM, the perturbation equations, Eqs.~\eqref{poissoneqlnaSHAL}--\eqref{eulereqlnaSHAL}, coincide with the Newtonian limit of general relativity, as also shown in Appendix~\ref{NPT}. This agreement is obtained from the sub-horizon condition $\epsilon\ll1$, without imposing an additional quasi-static assumption. The SHA equations obtained for the non-interacting scalar-field models likewise reduce to the same Newtonian-form system, but not in the case of the interacting model. Under the quasi-static approximation (QSA) in $\Lambda$CDM, only the continuity equation Eq. \eqref{continuityeqlnaQSAL} recovers its Newtonian form, while the Euler and Poisson equations Eqs. \eqref{eulereqlnaQSAL} and \eqref{poissoneqlnaQSAL} retain relativistic correction terms. The Newtonian form is recovered only when the sub-horizon condition is imposed in addition to the QSA. In the scalar field dark energy models, QSA has many more extra terms.

We first compare the full, SHA, and QSA evolution of the matter perturbations for the interacting scalar-field dark-energy model. The corresponding growth factor and growth rate for $(\gamma,\beta)=(0.04,0.001)$ are shown in Fig.~(\ref{Nonminimal_k}) for both quintessence and phantom dark energy at the representative wavenumbers considered in this work. The three solutions exhibit a clear scale dependence, with the differences between the approximate and full solutions becoming more pronounced towards smaller $k$. Over the range considered, the QSA remains closer to the full solution than the SHA.

The percentage errors $\mathcal{E}_{X}^{A}(a,k)$ of the SHA and QSA relative to the full solution for $\Lambda$CDM, the non-interacting scalar-field model, and the interacting models with $(\gamma,\beta)=(0.04,0.001)$ and $(0.06,0.001)$ are shown in Figs.~(\ref{SHA_k_error}) and~(\ref{QSA_k_error}) using Eq. \eqref{errorfromfullsoln}. For both the growth factor $D$ and growth rate $f$, the QSA generally exhibits smaller errors than the SHA. The errors increase towards smaller $k$, with the growth-rate errors showing a stronger variation over redshift. For the interacting models, increasing $\gamma$ also leads to a systematic increase in the approximation error.

The dependence of the SHA error on the interaction parameters is further examined in Fig.~(\ref{parameter_k_error}), which shows the error in $D$ at $z=0$ over the $(k,\gamma)$ parameter space. For $\gamma=0$, the error remains at the level of $\sim1$--$2\%$ over the considered range of $k$. The error remains comparably small for weak interactions but increases progressively with $\gamma$, exceeding the $5\%$ level in part of the parameter space even for modes well inside the horizon. The $5\%$ contour therefore denotes a $\gamma$- and $k$-dependent region within which the SHA remains within the adopted accuracy criterion.

It is also evident in the parameter-space results that there is very little distinction between the quintessence and phantom cases in error with respect to approximation. While the redshift evolution of the SHA and QSA errors in Figs.~(\ref{SHA_k_error}) and~(\ref{QSA_k_error}) is broadly similar for the two scalar-field natures, the error distributions at $z=0$ in Fig.~(\ref{parameter_k_error}) differ by only a few percent over the same $(k,\gamma)$ range, demonstrating that the approximation errors do not depend much on the nature of the scalar field.

We next compare the scalar-field models with $\Lambda$CDM using consistent full--full and SHA--SHA comparisons of the matter growth factor, as defined in Eq.~\eqref{percentage_deviation}. The results are shown in Fig.~(\ref{diff_lcdm}). The deviations from $\Lambda$CDM themselves exhibit only a weak dependence on $k$. In contrast, the difference between the full--full and SHA--SHA deviations shows a clearer scale dependence, increasing towards smaller $k$. This difference also increases with the interaction strength $\gamma$. The corresponding $\gamma$--$k$ dependence is examined in Fig.~(\ref{deviation}) over $\gamma\in[0,0.09]$ and $k\in[4\times10^{-3},10^{-1}],\mathrm{Mpc}^{-1}$ at $z=0$. For the non-interacting case, $\gamma=0$, the difference remains close to zero across the considered range. Increasing $\gamma$ produces a progressively larger difference.

We also investigated the impact of the approximations on the matter power spectrum at $z=0$. Fig. ~(\ref{percentage_error_P_full}) shows the percentage deviations of the full and SHA power spectra from the full $\Lambda$CDM power spectrum for the non-interacting model and the interacting models with $(\gamma,\beta)=(0.04,0.001)$ and $(0.06,0.001)$ using Eq. \eqref{powerspectrawrtlcdm}. The difference between the SHA and full deviations is also shown using Eq. \eqref{powerspectradeviation}. The discrepancy between the two treatments increases towards smaller $k$ and becomes more pronounced with increasing interaction strength. The wavenumbers at which this difference reaches $2\%$ and $5\%$ for the interacting model with $(\gamma,\beta)=(0.04,0.001)$ are identified in Fig. ~(\ref{diff}) for the interacting models. These threshold scales vary with $\gamma$, demonstrating that the range of wavenumbers over which the SHA satisfies a specified accuracy criterion is interaction-dependent.

Across the cases considered, the QSA provides a closer approximation to the full perturbation evolution than the SHA. The difference between the two approximations becomes increasingly evident at smaller $k$ and for larger interaction strength. The results therefore show that the accuracy of both approximations is scale-dependent, with the SHA exhibiting a stronger dependence on the interaction parameters.
\section{Discussion and conclusion}\label{Discussion}
\label{concl}

The results presented in this work establish the range of scales over which the SHA and QSA remain valid. Although the SHA condition is sufficient to recover the Newtonian fluid equations in \(\Lambda\)CDM and non-interacting scalar-field models, a dark-sector interaction introduces additional terms whose importance depends on the interaction strength. Consequently, the standard Newtonian equations are not fully recovered in the interacting case, even after applying the SHA. This distinction is relevant for numerical studies of structure formation, particularly \(N\)-body simulations, which commonly evolve perturbations using Newtonian fluid equations \cite{Silva:2024ift,Zhai:2026uwr}.

The SHA itself is reliable only for sufficiently weak interactions and becomes progressively less accurate as the interaction strength increases, even for modes well inside the horizon. The dependence on the interaction strength is particularly relevant for interpreting the validity of the approximation. In the non-interacting limit, for the full solutions, scalar-field perturbations become sufficiently suppressed on scales well inside the horizon such that the SHA reproduces the full evolution with a longer range of k values near the horizon. Once an interaction is introduced, however, the coupling between the dark matter and scalar-field sectors modifies the perturbation dynamics. The characteristic scale below which the SHA remains accurate is consequently no longer determined by the horizon scale alone. Instead, it depends on the interaction parameters and, to a much smaller extent, on the type of the scalar field, quintessence or phantom.
This behaviour indicates that the SHA cannot be characterized by a universal scale that is independent of the underlying dark energy model.

The comparison between SHA and QSA also has an important theoretical implication. The QSA retains more of the relativistic structure of the perturbation equations, whereas the SHA removes the expansion in powers of $\mathcal{H}/k$. The smaller QSA deviations found in our analysis therefore indicate that neglecting the time evolution of the gravitational potential is less restrictive than imposing the full sub-horizon reduction for the models and scales considered here. This suggests that, when relativistic accuracy is required but a fully dynamical treatment is computationally prohibitive, the QSA can provide a useful intermediate description. Nevertheless, its accuracy should also be assessed for each particular model, since the interaction-dependent behaviour found here demonstrates that neither approximation possesses a universal range of validity.

To isolate the performance of the SHA, we compare how much the interacting model deviates from $\Lambda$CDM under the full relativistic equations versus under the SHA framework. That is, we compare the relative departure of the full interacting solution from the full $\Lambda$CDM solution with that of the SHA interacting solution from the SHA $\Lambda$CDM baseline. Agreement between these deviations would indicate that the SHA adequately captures the underlying physical differences between the models. However, the increasing divergence between the two predictions at lower wavenumbers ($k$) and higher interaction strengths ($\gamma$) reveals that the approximation breaks down for stronger couplings. In this regime, the SHA fails to account for the complete departure from $\Lambda$CDM, which physically explains the growing approximation errors observed in interacting dark sector models.

These findings are particularly relevant for cosmological $N$-body and hydrodynamical simulations. Such simulations commonly evolve structure formation using Newtonian fluid equations because the computational cost of solving the complete relativistic perturbation system is substantially higher. This treatment is well motivated for standard $\Lambda$CDM on sufficiently small scales, and our results confirm its accuracy in the corresponding non-interacting limit. However, when the aim is to simulate interacting dark energy, the Newtonian treatment is no longer guaranteed to reproduce the full relativistic evolution at the same level of accuracy. The interaction strength must therefore be regarded as an additional parameter controlling the range of scales over which the Newtonian approximation can be safely employed.

This issue becomes increasingly important as simulation volumes continue to grow. Modern hydrodynamical simulations are being designed specifically to provide sufficiently large volumes for direct comparison with large-scale surveys. The FLAMINGO project, for example, includes a $(2.8,\mathrm{Gpc})^3$ hydrodynamical simulation, corresponding to a fundamental mode of approximately $k_{\rm min}\simeq2.24\times10^{-3},\mathrm{Mpc}^{-1}$, and is designed for applications to large-scale structure and galaxy-cluster surveys \cite{Schaye:2023jqv}. Such large volumes allow simulations to sample modes much closer to the horizon than smaller-box simulations. At the same time, the number of independent large-scale modes increases the statistical power of comparisons with observations. Consequently, an approximation error that was negligible in a smaller simulation can become relevant once the simulation is extended to larger scales.

The same consideration applies directly to observational analyses. Current and forthcoming galaxy surveys measure clustering over very large cosmological volumes, making the large-scale shape of the matter and galaxy power spectra increasingly important. In particular, DESI has already incorporated full-shape galaxy and quasar clustering information, including redshift-space distortions and the matter-radiation-equality scale, into cosmological parameter constraints \cite{DESI:2024jxi,DESI:2024hhd}. The statistical precision of such measurements makes theoretical consistency between the model used to interpret the observations and the approximation used to generate its predictions increasingly important. A scale-dependent theoretical error of several percent can therefore become comparable to, or larger than, the statistical precision of some measurements and may affect the inferred preference for parameters describing the dark sector.

The increasing precision of large-scale surveys also changes the significance of approximation errors. An error of $2$--$5\%$ may be relatively unimportant for an analysis dominated by smaller scales or by larger observational uncertainties. However, as simulations and surveys extend to larger volumes and increasingly exploit large-scale modes, the same approximation error can become a limiting theoretical systematic. If it is not included in the error budget or properly modelled, it can propagate into parameter estimation and potentially obscure or mimic the scale-dependent signatures expected from modified dark-sector physics. Thus, the accuracy requirement for theoretical predictions should be considered together with the statistical precision of the survey rather than treated as a fixed property of the approximation.

The $2\%$ and $5\%$ thresholds considered in this work consequently provide a useful practical way of connecting the theoretical approximation to simulation and observational requirements. For a specified interaction model, the limiting wavenumber at a chosen accuracy determines the largest box size for which the SHA remains within that accuracy, through $L_{\max}=2\pi/k_{\rm lim}$. Conversely, for a simulation of a given volume, one can determine whether the modes entering the analysis remain within the validated range of the approximation. This criterion is particularly useful when simulations are designed to reproduce the scales probed by a specific survey: the required simulation volume should not be selected independently of the accuracy of the perturbative approximation used to evolve those modes.

An important consequence is that increasing the simulation volume does not automatically improve the theoretical prediction. A larger box provides access to additional large-scale modes and reduces sample variance, but it also extends the calculation into a regime where the SHA error can become larger. Thus, the gain in statistical information from larger volumes must be accompanied by a corresponding control of theoretical systematics. This becomes increasingly relevant for next-generation surveys, for which statistical uncertainties are expected to decrease and theoretical modelling errors can become one of the dominant limitations.

Our analysis therefore suggests that the use of the SHA in interacting-dark-energy simulations should be viewed as a model-, scale-, and accuracy-dependent approximation rather than as a universal Newtonian prescription. For weak interactions and sufficiently large $k$, the SHA can provide a computationally efficient description of the perturbation evolution. For stronger interactions or larger scales, however, the approximation should first be benchmarked against the full relativistic equations. In practical applications, this validation can be performed once for a given interaction model and parameter range and then used to identify the region of $(k,\gamma,\beta)$ space in which Newtonian simulations remain reliable.

There are several directions in which this analysis can be extended. First, the present study focuses on linear perturbations and therefore provides the appropriate benchmark for identifying the scale at which the relativistic approximation begins to fail, but it does not by itself quantify the propagation of this error into the nonlinear regime. A natural next step is to incorporate the interaction consistently into nonlinear $N$-body or hydrodynamical simulations and determine whether the scale-dependent differences found here persist after nonlinear structure formation. Second, the impact of the approximation should be propagated directly into observable quantities such as galaxy clustering, weak gravitational lensing, and redshift-space distortions, where the relevant theoretical accuracy requirements can be compared with those of current and forthcoming surveys. Finally, extending the analysis over a broader range of interaction parameters and scalar-field potentials would allow the construction of a more general validity map for relativistic-to-Newtonian reductions in interacting dark-energy cosmologies.

Overall, the main implication of this work is that the validity of the SHA should be established relative to the scales and precision of the intended application. As cosmological simulations and observational surveys move towards larger volumes and increasingly precise measurements, approximation errors that were previously negligible can become relevant theoretical systematics. Explicitly benchmarking the approximation against the full perturbation dynamics therefore provides an essential consistency check before using interacting-dark-energy models for precision large-scale-structure predictions.

\acknowledgments
S. N-G would like to thank IIT Kanpur for the Initiation Grant (SSA/2022232) and the Department of Science and Technology, ANRF (previously Science and Engineering Research Board) for the MATRICS grant (No.MTR/2023/001376 ). 

\appendix

\section{Energy momentum conservation of the effective matter degree of freedom}
\label{emcon}

We now derive the background and linear perturbation energy-conservation
equations for the algebraically coupled dark sector. This derivation is
important because, although the interaction contribution can be absorbed into
an effective matter energy density, the resulting effective matter sector is
not separately conserved.

For the interaction considered in this work,
\begin{equation}
    \rho_{\rm int}
    =
    \gamma \rho_m e^{-\beta\phi},
    \qquad
    P_{\rm int}=0,
    \label{eq:rho_int_def_corr}
\end{equation}
where we have already imposed $\alpha=1$. The effective matter density is
therefore
\begin{equation}
    \rho_m^{\rm eff}
    =
    \rho_m+\rho_{\rm int}
    =
    \rho_m\left(1+\gamma e^{-\beta\phi}\right).
    \label{eq:rho_eff_def_corr}
\end{equation}
The derivative of the interaction density with respect to the scalar field is
\begin{equation}
    \rho_{{\rm int},\phi}
    \equiv
    \frac{\partial\rho_{\rm int}}{\partial\phi}
    =
    -\beta\rho_{\rm int}.
    \label{eq:rho_int_phi_corr}
\end{equation}

The scalar-field and effective-matter energy--momentum tensors obey
\begin{equation}
    \nabla_\mu T^{\mu}{}_{\nu(\phi)}
    =
    Q_\nu,
    \qquad
    \nabla_\mu T^{\mu}{}_{\nu(m)}^{\rm eff}
    =
    -Q_\nu,
    \label{eq:exchange_covariant_corr}
\end{equation}
where the interaction current is
\begin{equation}
    Q_\nu
    =
    \rho_{{\rm int},\phi}\,
    \partial_\nu\phi.
    \label{eq:Qnu_corr}
\end{equation}
Consequently,
\begin{equation}
    \nabla_\mu
    \left(
        T^{\mu}{}_{\nu(m)}^{\rm eff}
        +
        T^{\mu}{}_{\nu(\phi)}
    \right)
    =0,
    \label{eq:total_cons_corr}
\end{equation}
as required by the Bianchi identity.


\subsubsection{Background energy conservation}
\label{subsubsec:background_cons_corr}

For a spatially homogeneous scalar field,
\begin{equation}
    \phi=\bar\phi(\tau),
\end{equation}
the temporal component of Eq.~\eqref{eq:exchange_covariant_corr} gives
the effective-matter background energy equation,
\begin{equation}
    \bar\rho_m^{{\rm eff}\,\prime}
    +
    3\mathcal H\bar\rho_m^{\rm eff}
    =
    \bar\rho_{{\rm int},\phi}\,
    \bar\phi'.
    \label{eq:rhoeff_background_general}
\end{equation}
Using Eq.~\eqref{eq:rho_int_phi_corr}, this becomes
\begin{equation}
    \bar\rho_m^{{\rm eff}\,\prime}
    =
    -3\mathcal H\bar\rho_m^{\rm eff}
    -
    \beta\bar\rho_{\rm int}\bar\phi'\,.
    \label{eq:rhoeff_background_explicit}
\end{equation}
Thus, although $\bar\rho_m\propto a^{-3}$,
the effective matter density does not in general obey the dust scaling because
the interaction energy per matter particle depends on the evolving scalar
field.

The scalar-field action gives the interacting Klein--Gordon equation
\begin{equation}
    \epsilon_d \Box\phi
    -
    V_{,\phi}
    -
    \rho_{{\rm int},\phi}
    =0.
    \label{eq:KG_covariant_corr}
\end{equation}
For the homogeneous FLRW background this becomes
\begin{equation}
    \epsilon_d
    \left(
        \bar\phi''
        +
        2\mathcal H\bar\phi'
    \right)
    +
    a^2
    \left(
        V_{,\phi}
        +
        \bar\rho_{{\rm int},\phi}
    \right)
    =0.
    \label{eq:KG_background_general}
\end{equation}
Using Eq.~\eqref{eq:rho_int_phi_corr},
\begin{equation}
    \epsilon_d
    \left(
        \bar\phi''
        +
        2\mathcal H\bar\phi'
    \right)
    +
    a^2
    \left(
        V_{,\phi}
        -
        \beta\bar\rho_{\rm int}
    \right)
    =0\,.
    \label{eq:KG_background_explicit}
\end{equation}

Introducing $N\equiv\ln a$, with
$d/d\tau=\mathcal H\,d/dN$, Eq.~\eqref{eq:KG_background_explicit}
can equivalently be written as
\begin{equation}
    \frac{d^2\bar\phi}{dN^2}
    =
    -
    \left(
        2+
        \frac{d\ln\mathcal H}{dN}
    \right)
    \frac{d\bar\phi}{dN}
    -
    \frac{\epsilon_d a^2}{\mathcal H^2}
    \left(
        V_{,\phi}
        -
        \beta\bar\rho_{\rm int}
    \right)\,.
    \label{eq:KG_background_N_corr}
\end{equation}

\subsubsection{Linear perturbation of the effective matter energy equation}
\label{subsubsec:perturbed_energy_corr}

We decompose
\begin{equation}
    \phi
    =
    \bar\phi+\delta\phi,
    \qquad
    \rho_m
    =
    \bar\rho_m+\delta\rho_m.
\end{equation}
The bare matter density contrast is defined as
\begin{equation}
    \delta_m
    \equiv
    \frac{\delta\rho_m}{\bar\rho_m}.
    \label{eq:deltam_def_corr}
\end{equation}
Perturbing Eq.~\eqref{eq:rho_int_def_corr} gives
\begin{align}
    \delta\rho_{\rm int}
    &=
    \gamma e^{-\beta\bar\phi}\delta\rho_m
    -
    \beta\gamma\bar\rho_m
    e^{-\beta\bar\phi}\delta\phi =
    \bar\rho_{\rm int}
    \left(
        \delta_m-\beta\delta\phi
    \right).
    \label{eq:delta_rhoint_corr}
\end{align}
Hence the perturbation of the effective matter density is
\begin{align}
    \delta\rho_m^{\rm eff}
    &=
    \delta\rho_m+\delta\rho_{\rm int} =
    \bar\rho_m^{\rm eff}\delta_m
    -
    \beta\bar\rho_{\rm int}\delta\phi.
    \label{eq:delta_rhoeff_corr}
\end{align}
Defining
\begin{equation}
    \mathcal{C}
    \equiv
    \frac{\beta\bar\rho_{\rm int}}
         {\bar\rho_m^{\rm eff}},
    \label{eq:C_def_corr}
\end{equation}
one may equivalently write
\begin{equation}
    \delta_m^{\rm eff}
    \equiv
    \frac{\delta\rho_m^{\rm eff}}
         {\bar\rho_m^{\rm eff}}
    =
    \delta_m-\mathcal{C}\delta\phi.
    \label{eq:delta_eff_corr}
\end{equation}
The temporal component of the effective matter conservation law is
\begin{equation}
    \nabla_\mu
    T^{\mu}{}_{0(m)}^{\rm eff}
    =
    -Q_0.
    \label{eq:eff_temporal_cons_corr}
\end{equation}
At first order its left-hand side is
\begin{equation}
    \delta
    \left(
        \nabla_\mu
        T^{\mu}{}_{0(m)}^{\rm eff}
    \right)
    =
    -\delta\rho_m^{{\rm eff}\,\prime}
    -3\mathcal H\delta\rho_m^{\rm eff}
    +3\bar\rho_m^{\rm eff}\Phi'
    -\bar\rho_m^{\rm eff}\theta_m.
    \label{eq:eff_div_temporal_corr}
\end{equation}
From Eq.~\eqref{eq:Qnu_corr},
\begin{align}
    \delta Q_0
    &=
    -\beta\bar\rho_{\rm int}\delta\phi'
    -
    \beta\bar\phi'\delta\rho_{\rm int}=
    -\beta\bar\rho_{\rm int}\delta\phi'
    -
    \beta\bar\rho_{\rm int}\bar\phi'
    \left(
        \delta_m-\beta\delta\phi
    \right),
    \label{eq:deltaQ0_corr}
\end{align}
where Eq.~\eqref{eq:delta_rhoint_corr} has been used.
The perturbed effective-matter energy equation is consequently
\begin{align}
    &\delta\rho_m^{{\rm eff}\,\prime}
    +3\mathcal H\delta\rho_m^{\rm eff}
    -3\bar\rho_m^{\rm eff}\Phi'
    +\bar\rho_m^{\rm eff}\theta_m
    =
    -\beta\bar\rho_{\rm int}\delta\phi'
    -
    \beta\bar\rho_{\rm int}\bar\phi'
    \left(
        \delta_m-\beta\delta\phi
    \right).
    \label{eq:delta_rhoeff_energy_corr}
\end{align}
Equation~\eqref{eq:delta_rhoeff_energy_corr} is the correct perturbed
energy-conservation equation for the \emph{effective} matter density.

It
contains explicit interaction terms and therefore should not be replaced by
the conservation equation of an independently conserved dust component.
We can now demonstrate that the bare-CDM density contrast nevertheless obeys
the standard continuity equation. Differentiating
Eq.~\eqref{eq:delta_rhoeff_corr} gives
\begin{align}
    \delta\rho_m^{{\rm eff}\,\prime}
    &=
    \bar\rho_m^{{\rm eff}\,\prime}\delta_m
    +
    \bar\rho_m^{\rm eff}\delta_m'
    -
    \beta\bar\rho_{\rm int}'\delta\phi
    -
    \beta\bar\rho_{\rm int}\delta\phi'.
    \label{eq:delta_rhoeff_prime_corr}
\end{align}
Using
Eq.~\eqref{eq:rhoeff_background_explicit},
Eq.~\eqref{eq:delta_rhoint_corr}, and the corresponding background evolution
of $\bar\rho_{\rm int}$, all interaction-dependent terms appearing in
Eq.~\eqref{eq:delta_rhoeff_energy_corr} cancel. One is left with
\begin{equation}
    \bar\rho_m^{\rm eff}
    \left(
        \delta_m'
        +
        \theta_m
        -
        3\Phi'
    \right)
    =0.
    \label{eq:bare_continuity_intermediate}
\end{equation}
Since $\bar\rho_m^{\rm eff}\neq0$, the bare-CDM continuity equation is
\begin{equation}
    \delta_m'
    +
    \theta_m
    -
    3\Phi'
    =0\,.
    \label{eq:bare_continuity_corr}
\end{equation}
Equivalently, in terms of $N=\ln a$,
\begin{equation}
    \frac{d\delta_m}{dN}
    =
    -\frac{\theta_m}{\mathcal H}
    +
    3\frac{d\Phi}{dN}.
    \label{eq:bare_continuity_N_corr}
\end{equation}
This result has a clear interpretation. The effective matter energy density is
not separately conserved because it contains the scalar-dependent interaction
energy. The bare CDM particle number, on the other hand, remains conserved.
Consequently, the interaction does not appear as a direct particle-production
term in Eq.~\eqref{eq:bare_continuity_corr}. Its effect on matter clustering
enters through the momentum equation, where the transverse part of the
interaction current generates the fifth force.

\section{Newtonian Perturbation theory}\label{NPT}

Considering a pressure-less fluid with mass density $\rho$ and velocity $\mathbf{u}$. The Newtonian evolution of this self-gravitating fluid is governed by the continuity, Euler, and Poisson equations \cite{Baumann:2022mni}. The continuity equation expresses conservation of mass,
\begin{equation}
    \frac{\partial \rho}{\partial t}+ \nabla \cdot\left(\rho \mathbf{u}\right)
=0,
\end{equation}
the Euler equation describes momentum conservation,
\begin{equation}
    \frac{\partial \mathbf{u}}{\partial t}
+
\left(\mathbf{u}\cdot\nabla\right)\mathbf{u}
=
-\nabla\Phi,
\end{equation}
where $\Phi$ is the gravitational potential. This potential is described by the matter density through Poisson equation, 

\begin{equation}
\nabla^2\Phi = 4\pi G\rho.
\end{equation}

Now, introducing perturbations around the background, we get,
\begin{eqnarray}
    \rho=\bar{\rho}[1+\delta],\quad \mathbf{u}=Ha\mathbf{x}+\mathbf{v}.
\end{eqnarray}
where $\bar{\rho}$ is the homogeneous background density, $\delta$ is the matter density contrast, $\mathbf{v}$ is the peculiar velocity and $Ha\mathbf{x}$ is the hubble flow.
Linearizing the Newtonian fluid equations and retaining only first-order perturbations gives
\begin{align}
\dot{\delta}
+
\frac{\nabla \cdot \mathbf{v}}{a}
&=0,
\\[6pt]
\dot{\mathbf{v}}
+
H\mathbf{v}
&=
-\frac{\nabla\Phi}{a},
\\[6pt]
\nabla^2\Phi
&=
4\pi G a^{2}\bar{\rho}\delta.
\end{align}
Here, overdot denotes the derivative with respect to cosmic time $t$. These three equations govern the evolution of linear density perturbations in Newtonian theory. Traditionally, using SHA and QSA conditions together in $\Lambda$CDM model, one can get back these same equations from the corresponding relativistic equations.

\bibliographystyle{unsrturl} 
\bibliography{reference,modgrav}

\end{document}